\documentclass[%
reprint,
superscriptaddress,
amsmath,amssymb,
aps,
]{revtex4-1}
\usepackage{ragged2e} 
\usepackage{booktabs}
\usepackage{multirow}
\usepackage{rotating}
\usepackage{subcaption}

\usepackage{float}
\usepackage{caption}
\usepackage{bm} 
\usepackage{graphicx}
\usepackage{dcolumn}
\usepackage{bm}
\usepackage{hyperref}

\begin{document}
	
	\preprint{APS/123-QED}
	
	\title{Quasinormal Modes, Partial Transmission Probabilities, and Hod's Conjecture of Charged Black Holes in Perfect Fluid Dark Matter within Kalb–Ramond Gravity}
	\author{Zongyuan Qin}
	\affiliation{%
		College of Physics,Guizhou University,Guiyang,550025,China
	}%
	
	\author{Zheng-Wen Long}%
	\email{zwlong@gzu.edu.cn (corresponding author)}
	\affiliation{%
		College of Physics,Guizhou University,Guiyang,550025,China
	}%

	
	\begin{abstract}
		~~Within the Kalb--Ramond field-induced Lorentz-violating gravity, we investigate the perturbation dynamics and quasinormal mode (QNM) spectra of a charged black hole immersed in perfect fluid dark matter. The background spacetime is characterized by the Lorentz-violating parameter $\tau$, the electric charge $Q/M$, and the dark matter parameter $\lambda/M$. Owing to the non-zero vacuum expectation value of the KR field, the metric function approaches $1/(1-\tau)$ at infinity instead of $1$, rendering the spacetime non-asymptotically flat. Under the test-field approximation, the equations of motion for three types of perturbations (scalar, electromagnetic, and axial gravitational) are reduced to Schr\"{o}dinger-type radial equations with single-peak effective potentials. The QNM frequencies are cross-validated using the sixth-order WKB method and the time-domain Prony method, while the partial transmission probabilities are computed within the WKB scattering framework. Parameter scans show that $\tau$ exerts the strongest influence on the QNM spectrum, followed by $\lambda/M$, with $Q/M$ having the weakest effect; the same ordering holds for the partial transmission probabilities. Under the same parameters, the QNM frequencies of the three perturbations follow the order scalar $>$ electromagnetic $>$ gravitational, whereas the transmission probabilities exhibit the reverse order, reflecting the different impact of the effective potential height on the oscillation frequency and on the transmission probability. The Hod conjecture $|\Im(\omega)| \le \pi T_H$ holds throughout the parameter range examined.
	\end{abstract}

	\maketitle
	
	
\section{Introduction}
\label{sec:1}
	General relativity has withstood numerous experimental and observational tests since its inception\cite{LIGOScientific:2016lio,Schiff:1960gh,Shapiro:1968zza,Franchini:2025csk,Hojjati:2011xd,Shapiro:1964uw,LIGOScientific:2021sio,LIGOScientific:2018dkp}, maintaining remarkable accuracy. Over the past decade, breakthroughs in gravitational wave astronomy and black hole imaging have pushed the study of strong gravitational fields to unprecedented precision. Since the first direct detection of gravitational waves by LIGO in 2015\cite{LIGOScientific:2016aoc}, the perturbation dynamics of black holes---in particular, quasinormal modes (QNMs)---have become a central tool for testing strong-field gravitational theories\cite{Konoplya:2011qq,Cardoso:2016rao,Nollert:1993zz,Andersson:1996xw}. QNMs are damped oscillatory modes excited when a black hole is perturbed, independent of the specific form of the initial disturbance. The QNM spectrum is therefore regarded as a ``characteristic fingerprint'' of a black hole, providing a unique avenue for probing black hole properties and discriminating among gravitational theories\cite{Liu:2021xfb,Konoplya:2011qq,Dreyer:2002vy,Cardoso:2008bp,Cardoso:2001bb,Kunstatter:2002pj,Cardoso:2017soq,Ching:1998mxl}.
	
	Beyond gravitational theory itself, dark matter\cite{Ferrer:2017xwm,Nampalliwar:2021tyz,Xu:2021dkv,Turner:1984nf,deSwart:2017heh,Spergel:2015noa,Hu:2000ke,Bode:2000gq,Spergel:1999mh} constitutes one of the foremost challenges in contemporary physics. Observations such as galactic rotation curves and gravitational lensing point to the existence of vast amounts of invisible matter, i.e., dark matter. To describe the behavior of black holes in a dark matter environment, Li and Yang proposed modeling dark matter as an anisotropic fluid\cite{Li:2012zx}. Although this model is often referred to as ``perfect fluid dark matter'' (PFDM) in the literature, its equation of state, $p_r = -\rho$ and $p_t = \rho/2$, indicates that the fluid is in fact anisotropic. This model has attracted considerable attention because it naturally accounts for gravitational effects on galactic scales\cite{Potapov:2016obe,Haroon:2018ryd,Narzilloev:2020qtd,Atamurotov:2021hoq,Rizwan:2018rgs,Huang:2026cjn,Zhu:2024lht}.
	
	On the other hand, Lorentz symmetry, as one of the cornerstones of modern physics, has been verified by high-precision experiments over an extremely wide energy range. Nevertheless, many frontier theories, such as string theory and loop quantum gravity, predict that Lorentz symmetry may be violated under certain conditions\cite{Kostelecky:1988zi,Alfaro:2001rb,Horava:2009uw,Carroll:2001ws,Jacobson:2000xp,Dubovsky:2004ud,Bengochea:2008gz,Cohen:2006ky,Kostelecky:2003fs}. Among various modification schemes, the Kalb--Ramond (KR) field has attracted significant interest owing to its unique theoretical origins\cite{Junior:2024ety,Kostelecky:1989jw,Maluf:2018jwc,Lessa:2019bgi,Rey:1989ti,Belich:2010xj,Mavromatos:2016mnj,Kumar:2020hgm,Du:2013bx,Chatterjee:2005cr,Junior:2024vdk}. The KR field is described by a rank-two antisymmetric tensor field, and when it acquires a non-zero vacuum expectation value, it leads to spontaneous, rather than explicit, breaking of local Lorentz symmetry. This provides a theoretical framework for testing the applicability of Lorentz invariance in the strong-field regime. In recent years, black hole solutions with non-minimal coupling between the KR field and gravity have been successively proposed, and their thermodynamic properties, shadows, lensing effects, etc., have been extensively studied\cite{Baruah:2025ifh,Junior:2024vdk,Filho:2023ycx}.
	
	Against this background, a considerable number of works have investigated black holes in Lorentz-violating gravity induced by the KR field or in a PFDM environment. For example, Deng et al.\cite{Deng:2025atg} studied QNMs of slowly rotating KR black holes; Yang et al.\cite{Yang:2023wtu} analyzed the thermodynamic properties of static, spherically symmetric black holes in the presence of a KR field; Guo et al.\cite{Guo:2023nkd} investigated the QNMs and greybody factors of Lorentz-violating black holes in KR gravity; Baruah et al.\cite{Baruah:2025ifh} explored the influence of a global monopole charge on black holes in KR gravity; and Rahmatov et al.\cite{Rahmatov:2025gpk} examined the gravitational lensing effects of black holes surrounded by PFDM in KR gravity. In addition, the QNMs and greybody factors of various black holes in a PFDM environment have been widely discussed.\cite{Jusufi:2019ltj,Tovar:2025apz,Silva:2026qlr,Das:2023ess}
	
	However, a systematic study that simultaneously computes the QNM spectra and partial transmission probabilities for scalar, electromagnetic, and axial gravitational perturbations on the charged KR--PFDM metric given by Ahmed et al.\cite{Ahmed:2026doo} has not yet been carried out. Furthermore, the validity of Hod's conjecture in this model has not been tested. The present work aims to fill this gap.
	
	Specifically, based on the charged KR--PFDM black hole solution proposed by Ahmed et al.\cite{Ahmed:2026doo}, we systematically investigate the QNM spectra, partial transmission probabilities, and the validity of Hod's conjecture for scalar, electromagnetic, and axial gravitational perturbations. The results show that the Lorentz-violating parameter $\tau$ has the most significant impact on the QNM spectrum, followed by the dark matter parameter $\lambda/M$, while the charge $Q/M$ exerts the weakest influence; the oscillation frequency of the gravitational perturbation is notably less affected by $\tau$ than those of the scalar and electromagnetic perturbations. Through these analyses, we construct a complete physical picture spanning from the QNM spectrum to the partial transmission probabilities and finally to Hod's conjecture, revealing how each parameter affects the perturbation properties and thermodynamic behavior of the black hole.
	
	The paper is organized as follows. Sect.\ref{sec:2} introduces the charged KR--PFDM black hole solution and its basic properties. Sect.\ref{sec:3} derives the effective potentials for the three types of test-field perturbations, computes the QNM frequencies, and analyzes how the QNM spectrum varies with the parameters. Sect.\ref{sec:4} calculates the partial transmission probabilities and performs a self-consistency check with the QNM results. Sect.\ref{sec:5} examines the validity of Hod's conjecture in this black hole model. Sect.\ref{sec:6} summarizes the paper. Throughout this work, we adopt geometric units $G=c=1$ and the metric signature $(-,+,+,+)$.

\section{Brief review of CHARGED BH SURROUNDED BY PFDM IN KR-GRAVITY}
\label{sec:2}
	
We briefly review the background black hole solution adopted in this paper; detailed derivations can be found in Ref.\cite{Ahmed:2026doo}. This solution describes a static, spherically symmetric spacetime influenced simultaneously by a Kalb--Ramond (KR) field, an electromagnetic field, and perfect fluid dark matter (PFDM). The total action can be written as\cite{Duan:2023gng,Yang:2023wtu,Jumaniyozov:2025dyy}
	
	\begin{equation}
		S = S_{\rm GR} + S_{\rm KR} + S_{\rm EM} + S_{\rm PFDM}
		\label{eq:1}
	\end{equation}
	
	where $S_{\rm GR}$ is the Einstein--Hilbert term, $S_{\rm KR}$ denotes the contribution of the KR field, $S_{\rm EM}$ is the electromagnetic action, and $S_{\rm PFDM}$ accounts for the PFDM. The KR field is described by a second-rank antisymmetric tensor \(B_{\mu\nu}\) whose non-vanishing vacuum expectation value \(\langle B_{\mu\nu}\rangle = b_{\mu\nu}\) spontaneously breaks local Lorentz symmetry. The electromagnetic field is governed by the vector potential $A_\mu$ and its field strength $F_{\mu\nu}$. The PFDM is modeled as an anisotropic fluid; in static spherically symmetric coordinates its energy--momentum tensor reads\cite{Jha:2025uie}
	\begin{equation}
		T_{\mu\nu}^{\rm PFDM} = {\rm diag}\bigl(\rho,\, p_r,\, p_t,\, p_t\bigr)
		\label{eq:2}
	\end{equation}
~~with the energy density $\rho = \lambda/(8\pi r^3)$, radial pressure $p_r = -\rho$, tangential pressure $p_t = \rho/2$, and the dark matter intensity parameter $\lambda$. This equation of state exhibits an anisotropic character ($p_r \neq p_t$).

~~The modified Einstein equations are obtained by varying the total action~\eqref{eq:1} with respect to the metric.
	\begin{equation}
		R_{\mu\nu} - \frac12 g_{\mu\nu} R = T_{\mu\nu}^{\rm KR} + T_{\mu\nu}^{\rm EM} + T_{\mu\nu}^{\rm PFDM}
		\label{eq:3}
	\end{equation}
	~~Assuming static spherical symmetry, the line element takes the form
	\begin{equation}
		ds^2 = -f(r)\,dt^2 + \frac{dr^2}{f(r)} + r^2\bigl(d\theta^2 + \sin^2\theta\,d\phi^2\bigr)
		\label{eq:4}
	\end{equation}
~~Combining the above equations, Ahmed et al.\cite{Ahmed:2026doo} obtained an exact black hole solution. The metric function reads
	\begin{equation}
		 f(r) = \frac{1}{1-\tau} - \frac{2M}{r} + \frac{Q^2}{(1-\tau)^2 r^2} + \frac{\lambda}{r}\ln\frac{r}{|\lambda|} 
		\label{eq:5}
	\end{equation}
	~~where $\tau$ is the dimensionless Lorentz-violating parameter, $M$ the black hole mass, $Q$ the electric charge parameter, and $\lambda$ the PFDM intensity parameter.
	~~The limiting behavior of this black hole solution clearly reflects the independent contributions of each physical parameter. When $Q = 0$, the metric reduces to a neutral black hole immersed in dark matter fluid under Lorentz-violating background\cite{Jumaniyozov:2025dyy,Jha:2025uie}. As $\lambda \to 0$, the dark matter effect disappears and the solution returns to the purely charged KR Lorentz-violating black hole\cite{Duan:2023gng}. For $\tau \to 0$, the Lorentz-violating effect decouples and the metric simplifies to a Reissner--Nordstr\"om black hole surrounded by PFDM\cite{Sadeghi:2024wib}. Finally, taking $\tau = Q = \lambda = 0$ naturally recovers the Schwarzschild solution.
	
	~~It is worth noting that at spatial infinity the metric function behaves as
	\begin{equation}
		\lim_{r\to\infty} f(r) = \frac{1}{1-\tau} \neq 1
		\label{eq:6}
	\end{equation}
	~~indicating that the spacetime is not asymptotically flat. This non-asymptotic flatness is a direct consequence of the KR field modifying the spacetime geometry.
	
	~~Based on the observational constraints from high-frequency QPOs analyzed via the MCMC method in Ref.\cite{Ahmed:2026doo}, the physically plausible ranges of the Lorentz-violating parameter $\tau/M$, the electric charge $Q/M$, and the dark matter parameter $\lambda$ are restricted to $\tau \in (0.05,\,0.20)$, $ Q/M \in (0.08,\,0.50)$, $ \lambda/M \in (0.15,\,0.45)$.
	
\begin{figure*}[htbp]
	\centering
	\renewcommand{\thesubfigure}{\roman{subfigure}}  
	\subcaptionbox{ $\tau=0.1,\lambda/M=0.2$}{
		\includegraphics[width=0.45\textwidth]{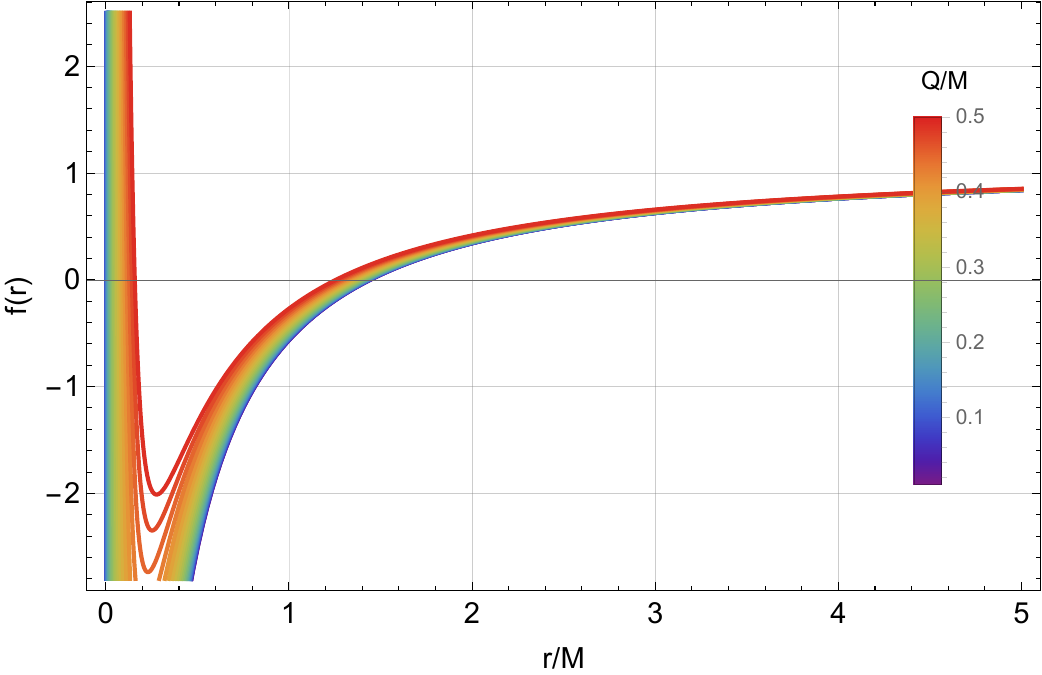} 
	}
	\hfill
	\subcaptionbox{ $Q/M=0.5,\tau=0.1$}{
		\includegraphics[width=0.45\textwidth]{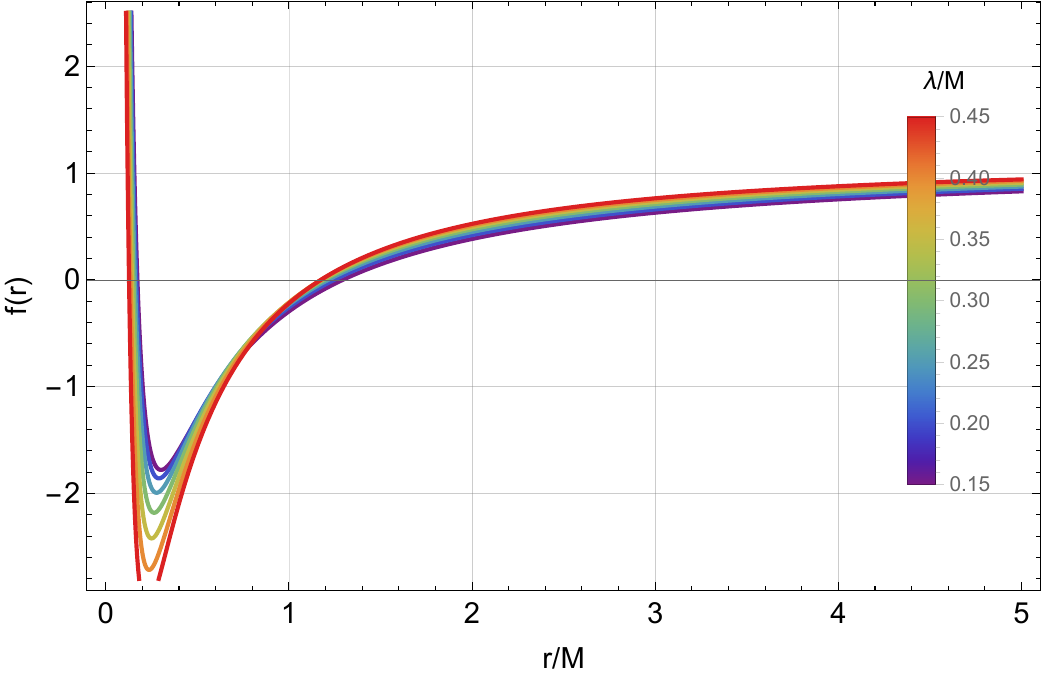} 
	}
	\hfill
	\subcaptionbox{ $\lambda/M=0.2,Q/M=0.5$}{
		\includegraphics[width=0.45\textwidth]{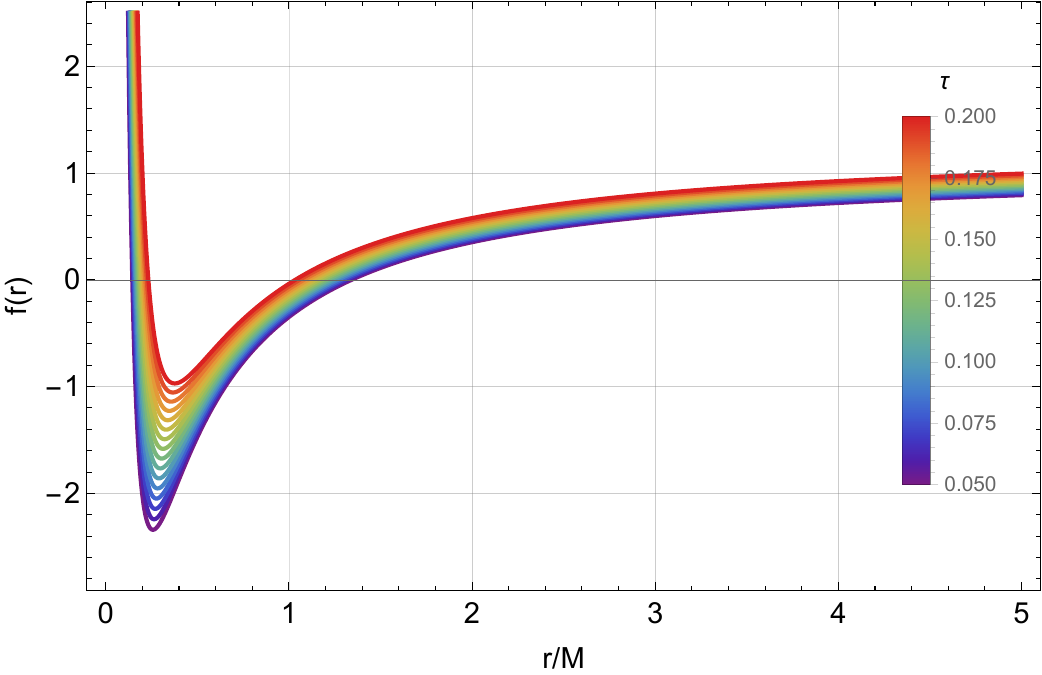} 
	}
	\caption{Variation of the metric function with the parameters $Q/M$, $\tau$, and $\lambda/M$ for a charged KR black hole in PFDM, with $M=1$  fixed.}
	\label{fig:1}
\end{figure*}

	 ~~Fig.\ref{fig:1} shows the metric function \(f(r)\) as a function of the radial coordinate \(r/M\) for a static, spherically symmetric black hole spacetime under the combined influence of a KR field, an electromagnetic field, and PFDM. The three subfigures, each obtained by fixing a different pair of parameters, systematically reveal the impact of the Lorentz-violating parameter $\tau$, the electric charge $Q/M$, and the PFDM parameter $\lambda/M$ on the spacetime geometry. A consistent trend is observed across all subfigures: increasing any of these three parameters always shrinks the event horizon radius $r_h$,  yet each parameter accomplishes this through a different mechanism. The parameter \(\tau\) modifies the constant term \(1/(1-\tau)\) in the metric function, thereby altering the asymptotic value at infinity and rendering the spacetime non-asymptotically flat; it generates a global correction to the spacetime geometry. In contrast, the contributions associated with \(Q/M\) and \(\lambda/M\) decay with increasing radial distance and mainly affect the metric in the vicinity of the horizon.  Furthermore, the three parameters exhibit markedly different far-field asymptotic behaviors. In subfigures (i) and (ii), the asymptotic value $f(\infty)=1/(1-\tau)$ is independent of the  parameters $Q/M$ and $\lambda/M$, respectively, so all curves converge as $r\to\infty$. In subfigure (iii), the asymptotic value itself increases with \(\tau\), and the curves at large radii do not coincide, clearly reflecting the global modification of the spacetime structure induced by the Lorentz-violating effect.
	
	\begin{figure*}[htbp]
		\centering
		\renewcommand{\thesubfigure}{\roman{subfigure}}  
		\subcaptionbox{ $Q/M=0.5,M=1$}{
			\includegraphics[width=0.45\textwidth]{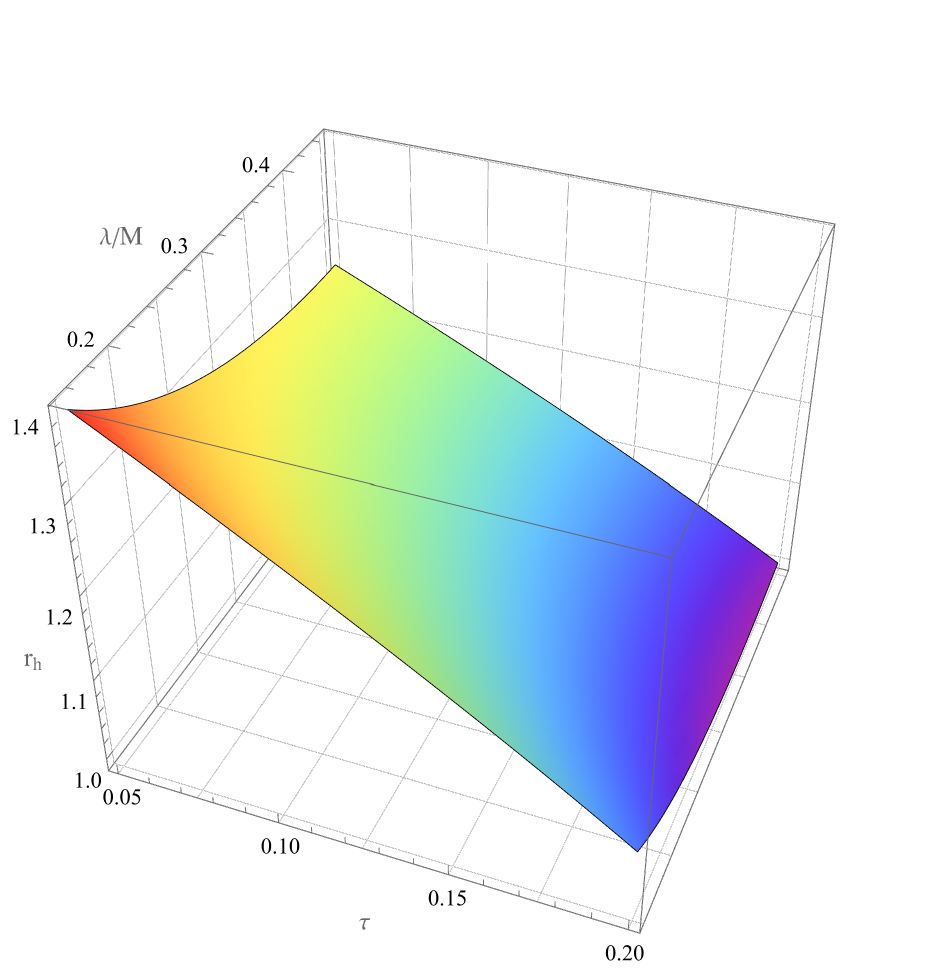} 
		}
		\hfill
		\subcaptionbox{ $\lambda/M=0.2,M=1$}{
			\includegraphics[width=0.45\textwidth]{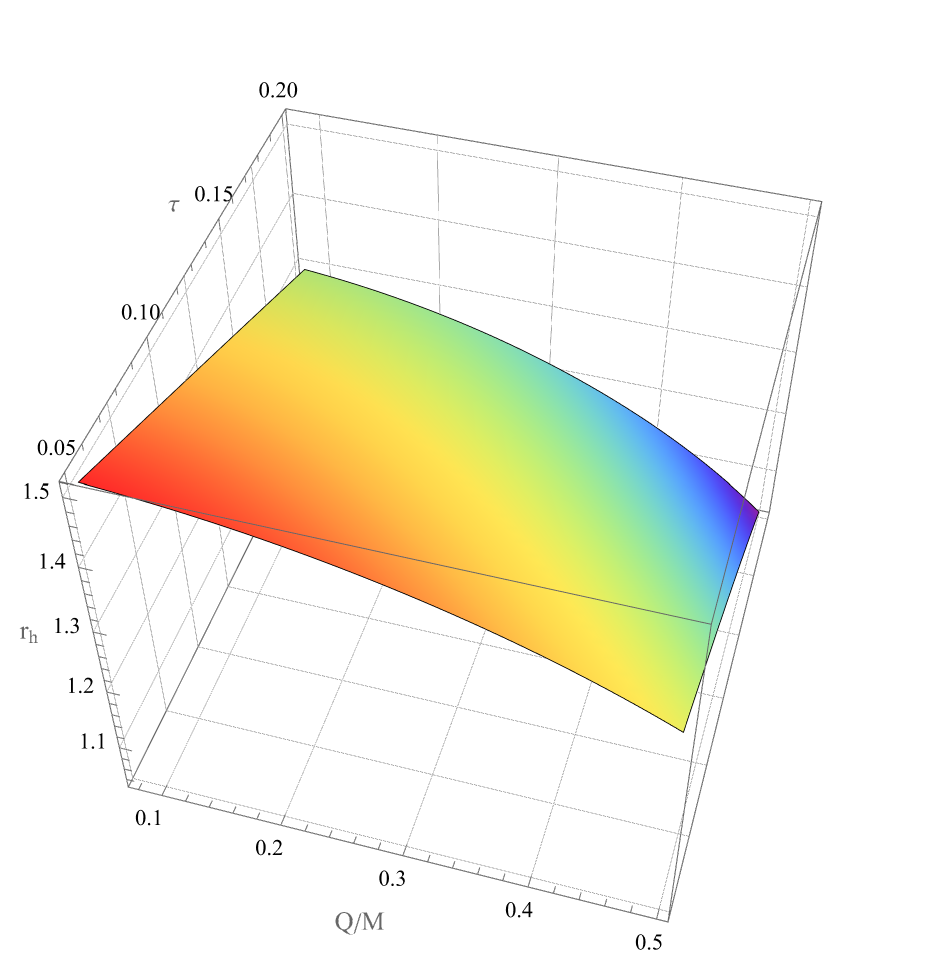} 
		}
		\hfill
		\subcaptionbox{ $\tau=0.1,M=1$}{
			\includegraphics[width=0.45\textwidth]{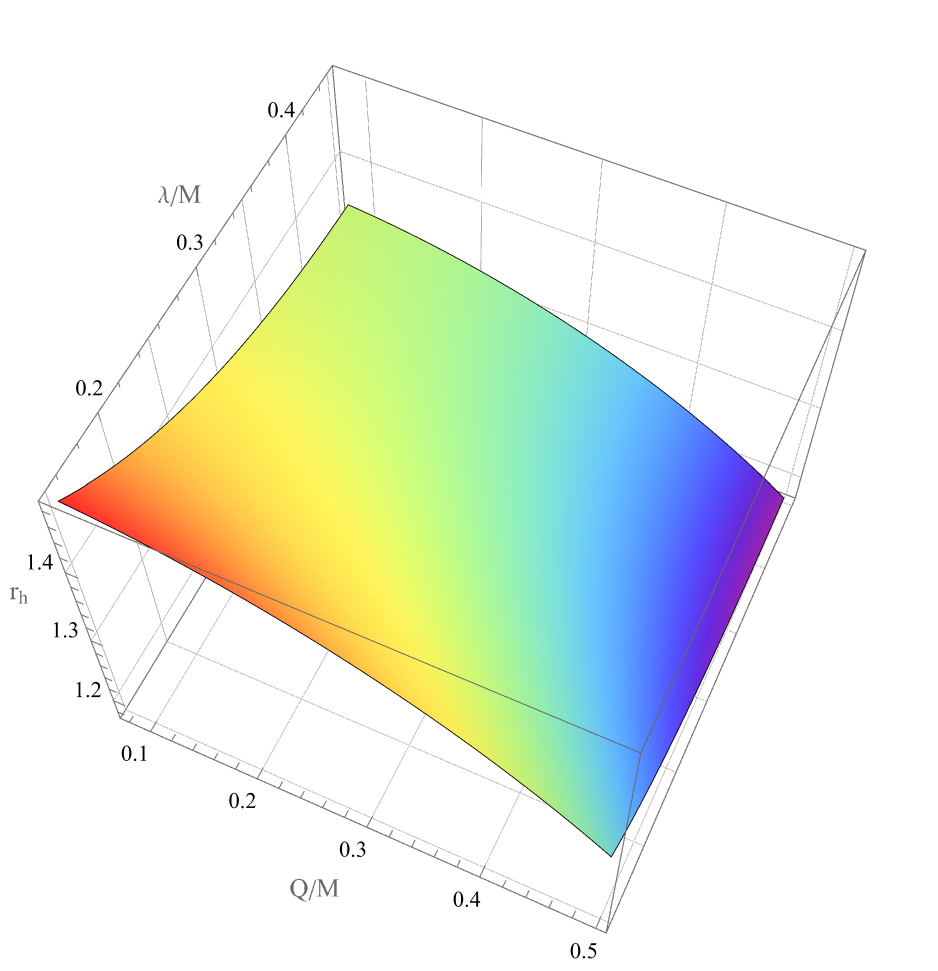} 
		}
		\caption{Investigation of the variation of the event horizon radius $r_h$ with the parameters $Q/M$, $\tau$, and $\lambda/M$ in a charged KR-PFDM spacetime.}
		\label{fig:2}
	\end{figure*}
	
	~~Fig.\ref{fig:2} shows the event horizon radius \(r_h\) of the charged KR--PFDM black hole as a function of the Lorentz-violating parameter \(\tau\), the charge parameter \(Q/M\), and the PFDM parameter \(\lambda/M\). By fixing $\tau=0.1$, $\lambda/M=0.2$, and $Q/M=0.5$ successively, we plot the three-dimensional surfaces of $r_h$ in the remaining two-parameter planes in panels (i)--(iii). Throughout the entire parameter region where an event horizon exists, $r_h$ exhibits a monotonically decreasing trend as $\tau$, $Q/M$, or $\lambda/M$ increases. This universal horizon contraction arises from the combined action of the correction terms in the metric function.

\section{PERTURBATION EQUATIONS and QUASINORMAL MODES}
\label{sec:3}
	This section investigates the dynamical behavior of the black hole under three different types of perturbations and the corresponding quasinormal mode (QNM) spectra. Specifically, we consider the propagation of a scalar field ($s=0$), an electromagnetic field ($s=1$), and axial gravitational perturbations ($s=2$) on this background spacetime.
	
	~~Given that in the perturbative limit the energy--momentum tensor of the perturbing field is far smaller than the total energy--momentum tensor of the background spacetime, and that the background KR and PFDM fields do not participate in the dynamical evolution of the perturbation, this work adopts the test-field approximation, studying only the propagation of perturbations on a fixed background metric, thereby avoiding the need to solve the coupled system of equations involving background-field perturbations\cite{Guo:2023nkd,Bouhmadi-Lopez:2020oia}.
	
	In a static spherically symmetric background, all three types of perturbations can be reduced, via separation of variables, to a unified Schr\"{o}dinger-type radial equation of the form
	\begin{equation}
		\frac{d^2\psi}{dr_*^2} + \bigl[\omega^2 - V(r)\bigr]\psi = 0 \label{eq:7}
	\end{equation}
	~~where $r_*$ is the tortoise coordinate and $V(r)$ is the effective potential for the corresponding perturbation type.
	Eq.\eqref{eq:7} satisfies the characteristic QNM boundary conditions
\begin{equation}
	\begin{aligned}
		\psi &\sim e^{-i\omega r_{*}} \quad (r_{*} \to -\infty)\\
		\psi &\sim e^{+i\omega r_{*}} \quad (r_{*} \to +\infty)
	\end{aligned}
	\label{eq:8}
\end{equation}
	~~The above boundary-value problem admits non-trivial solutions only for a discrete set of complex frequencies \(\omega = \omega_R + i\omega_I\).  In the following, we will derive the effective potentials for the three types of perturbations separately.

\renewcommand{\thesubsection}{\arabic{section}.\arabic{subsection}}

\subsection{The massless scalar field perturbations}
\label{sec:3.1}
The motion of a massless scalar field $\Phi$ in a curved spacetime is governed by the Klein--Gordon equation
\begin{equation}
	\frac{1}{\sqrt{-g}}\partial_{\mu}\bigl(\sqrt{-g}\,g^{\mu\nu}\partial_{\nu}\Phi\bigr)=0
	 \label{eq:9}
\end{equation}
~~For the static spherically symmetric metric\eqref{eq:5}, we decompose the scalar field in terms of spherical harmonics:
\begin{equation}
	\Phi(t,r,\theta,\phi)=\frac{1}{r}\sum_{l,m}\psi_{lm}(t,r)\,Y_{lm}(\theta,\phi)
	\label{eq:10}
\end{equation}
~~where $Y_{lm}(\theta,\phi)$ are the spherical harmonic functions, $l$ and $m$ are the angular and magnetic quantum numbers, respectively, and the angular quantum number satisfies $l\ge0$. Using the eigenvalue equation for the spherical harmonics, $\nabla_{\Omega}^{2}Y_{lm}=-l(l+1)Y_{lm}$, the angular part is separated. Assuming a time dependence of the form $e^{-i\omega t}$ for the radial function, i.e., $\psi_{lm}(t,r)=\psi_{lm}(r)e^{-i\omega t}$, and introducing the tortoise coordinate $r_{*}$ defined by $dr_{*}=\sqrt{\frac{g_{rr}}{|g_{tt}|}}\,dr$,
 the radial equation can be cast into a Schr\"{o}dinger-type form after separation of variables:
\begin{equation}
	\frac{d^{2}\psi}{dr_{*}^{2}}+\bigl[\omega^{2}-V_{s}(r)\bigr]\psi=0 
	\label{eq:11}
\end{equation}
The effective potential for the scalar field perturbation is given by
\begin{equation}
	V_{s}(r)=f(r)\left[\frac{l(l+1)}{r^{2}}+\frac{1}{r}\frac{df(r)}{dr}\right]
	\label{eq:12}
\end{equation}

This effective potential consists of the centrifugal barrier $\displaystyle f(r)\frac{l(l+1)}{r^{2}}$ and the curvature coupling term $\displaystyle \frac{f(r)f'(r)}{r}$. It is a universal result that holds rigorously for a scalar field in a static spherically symmetric background.
		\begin{figure*}[htbp]
			\centering
			\renewcommand{\thesubfigure}{\roman{subfigure}}  
			\subcaptionbox{$\tau=0.1$, $\lambda/M=0.2$}{
				\includegraphics[width=0.3\textwidth]{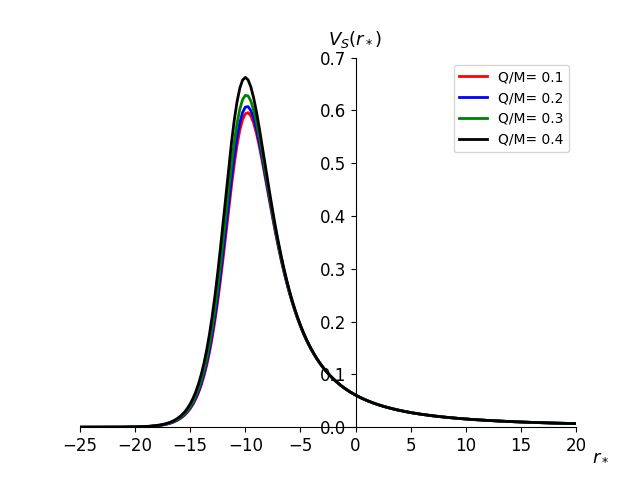}
			}
			\hfill
			\subcaptionbox{$\tau=0.1$, $Q/M=0.5$}{
				\includegraphics[width=0.3\textwidth]{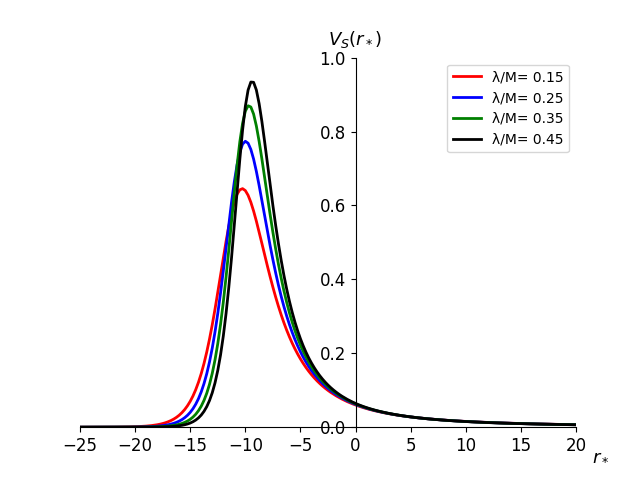}
			}
			\hfill
			\subcaptionbox{$\lambda/M=0.2$, $Q/M=0.5$}{
				\includegraphics[width=0.3\textwidth]{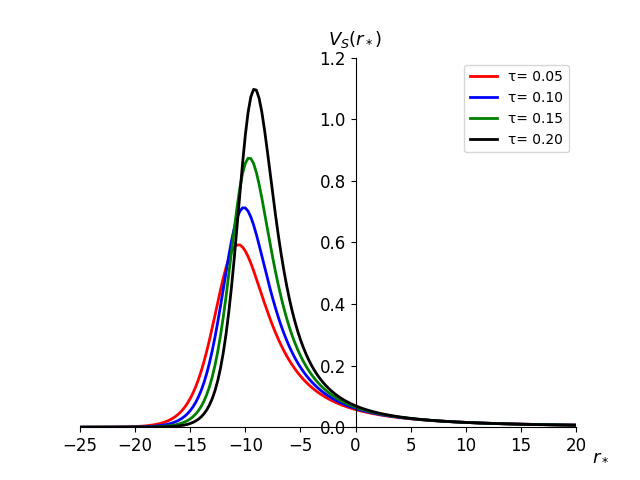}
			}
			\caption{Variation of the effective potential $V_s(r_*)$ with the tortoise coordinate $r_*$ for a charged KR black hole immersed in PFDM under a scalar field perturbation ($M=1$, $l=2$).}
			\label{fig:3}
		\end{figure*}
\subsection{The electromagnetic perturbations}
\label{sec:3.2}
	
	In a static spherically symmetric background, the spacetime possesses spatial reflection symmetry. Linear electromagnetic perturbations can be classified by parity into two independent modes: odd-parity (axial) and even-parity (polar). These two modes decouple from each other in the equations of motion, and their QNM spectra are completely degenerate. Therefore, we only study the odd-parity electromagnetic perturbations in this paper, and the results apply equally to the even-parity modes. For electromagnetic perturbations, within the test-field approximation, we neglect the non-minimal coupling between the electromagnetic perturbation and the KR field, and only consider the gravitational effect of the background metric on the electromagnetic propagation. In this case, the electromagnetic field satisfies the source-free Maxwell equations
	\begin{equation}
		\frac{1}{\sqrt{-g}}\partial_{\nu}\bigl(\sqrt{-g}\,F^{\mu\nu}\bigr)=0\qquad F_{\mu\nu}=\partial_{\mu}A_{\nu}-\partial_{\nu}A_{\mu} \label{eq:13}
	\end{equation}
	~~Adopting the transverse gauge $A_{t}=A_{r}=0$, the non-vanishing components of the four-potential for odd-parity perturbations can be expanded in terms of scalar spherical harmonics as\cite{Moncrief:1974gw}
	\begin{equation}
		A_{\theta}=\frac{a(r,t)}{\sin\theta}\frac{\partial Y_{lm}}{\partial\phi}\quad
		A_{\phi}=-a(r,t)\sin\theta\frac{\partial Y_{lm}}{\partial\theta} \label{eq:14}
	\end{equation}
	~~where $Y_{lm}(\theta,\phi)$ are the scalar spherical harmonics satisfying $\nabla_{\Omega}^{2}Y_{lm}=-l(l+1)Y_{lm}$, and the angular quantum number $l\ge1$. Substituting (14) into the Maxwell equations (13) and using the eigenvalue properties of the spherical harmonics to separate the angular part, we obtain the radial equation
	\begin{equation}
		\frac{\partial^{2}a}{\partial t^{2}}-f(r)\frac{\partial}{\partial r}\!\left[f(r)\frac{\partial a}{\partial r}\right]
		+f(r)\frac{l(l+1)}{r^{2}}a=0
\label{eq:15}
	\end{equation}
	~~Introducing the tortoise coordinate $\mathrm{d}r_{*}=\mathrm{d}r/f(r)$ and setting $\psi_{\mathrm{em}}(r,t)=a(r,t)$, Eq.~(15) can be cast into a standard Schr\"{o}dinger-type wave equation
	\begin{equation}
		\frac{\partial^{2}\psi_{\mathrm{em}}}{\partial t^{2}}-\frac{\partial^{2}\psi_{\mathrm{em}}}{\partial r_{*}^{2}}
		+V_{\mathrm{em}}(r)\psi_{\mathrm{em}}=0
	\label{eq:16}
	\end{equation}
	~~Taking $\psi_{\mathrm{em}}(r,t)=\varphi_{\mathrm{em}}(r)e^{-i\omega t}$ and substituting into Eq.~(16), we arrive at the radial equation
	\begin{equation}
		\frac{d^{2}\varphi_{\mathrm{em}}}{dr_{*}^{2}}+\bigl[\omega^{2}-V_{\mathrm{em}}(r)\bigr]\varphi_{\mathrm{em}}=0 
	\label{eq:17}
	\end{equation}
	~~The effective potential for the electromagnetic perturbation is given by
	\begin{equation}
		V_{\mathrm{em}}(r)=f(r)\frac{l(l+1)}{r^{2}}
		\label{eq:18}
	\end{equation}
	~~Compared with the effective potential for the scalar field, the electromagnetic effective potential does not contain the $f'(r)/r$ term, which is the characteristic form for a massless spin $s=1$ gauge field propagating in a static spherically symmetric spacetime. 
		\begin{figure*}[htbp]
		\centering
		\renewcommand{\thesubfigure}{\roman{subfigure}}  
		\subcaptionbox{$\tau=0.1$, $\lambda/M=0.2$}{
			\includegraphics[width=0.3\textwidth]{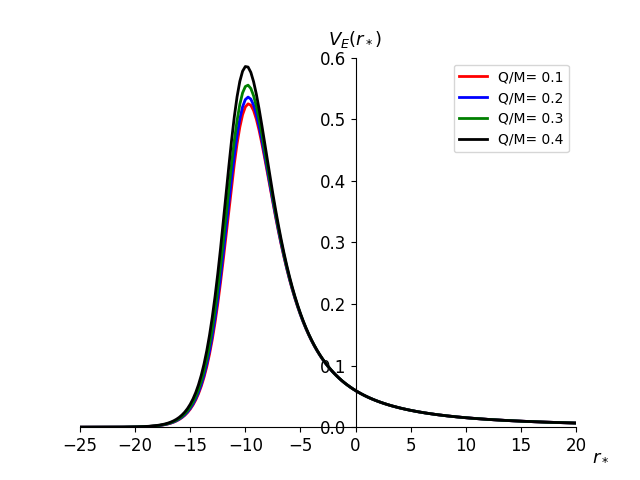}
		}
		\hfill
		\subcaptionbox{$\tau=0.1$, $Q/M=0.5$}{
			\includegraphics[width=0.3\textwidth]{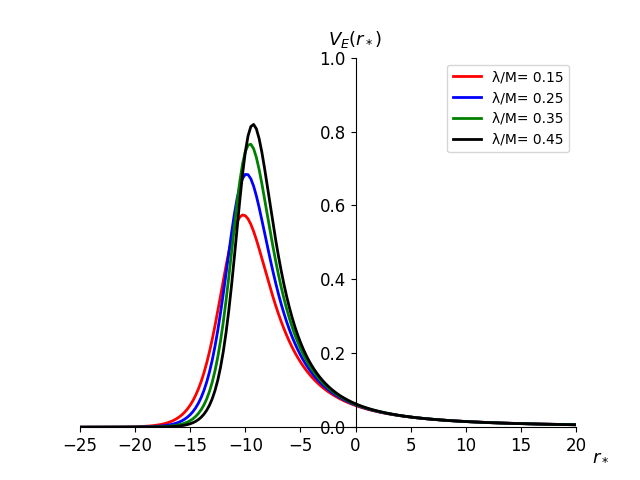}
		}
		\hfill
		\subcaptionbox{$\lambda/M=0.2$, $Q/M=0.5$}{
			\includegraphics[width=0.3\textwidth]{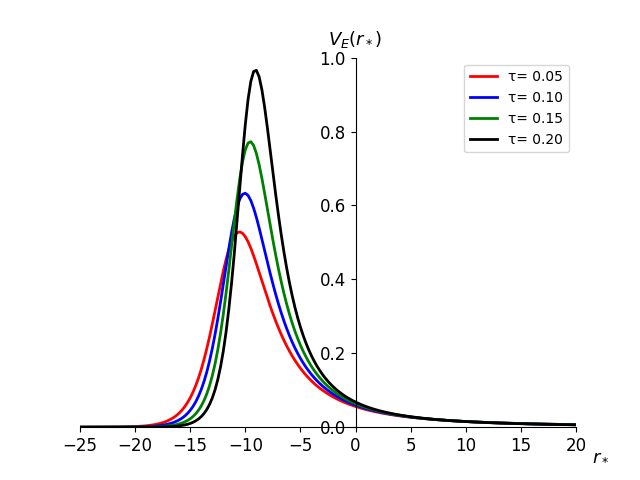}
		}
		\caption{Variation of the effective potential $V_E(r_*)$ with the tortoise coordinate $r_*$ for a charged KR black hole immersed in PFDM under a electromagnetic field perturbation ($M=1$, $l=2$).}
		\label{fig:4}
	\end{figure*}
	\subsection{The axial gravitational perturbations}
	\label{sec:3.3}
	In a static spherically symmetric background, linear gravitational perturbations can be classified by parity into axial (odd‑parity) and polar (even‑parity) independent modes\cite{Davis:1972ud}. We focus on axial gravitational perturbations in this work because their effective potential possesses a simpler analytic form. For axial perturbations, in the Regge--Wheeler gauge, the perturbed metric can be written as\cite{Regge:1957td}
\begin{equation}
	\begin{aligned}
		ds^{2} &= - f(r)dt^{2} + \frac{dr^{2}}{f(r)} + r^{2}\bigl(d\theta^{2}+\sin^{2}\theta d\phi^{2}\bigr)\\
		&\quad +2\sum_{l,m}\biggl[h_{0}(r)\sin\theta\frac{\partial Y_{lm}}{\partial\theta}\,dt d\phi
		+h_{1}(r)\sin\theta\frac{\partial Y_{lm}}{\partial\theta}\,dr d\phi\biggr]
	\end{aligned}
	\label{eq:19}
\end{equation}
	~~where \(h_{0}(r)\) and \(h_{1}(r)\) are two independent functions describing the axial perturbations.
	
	~~Under the test‑field approximation adopted in this paper, the first‑order perturbations of the background KR field and the PFDM field are set to zero, meaning that these background fields do not participate in the dynamical evolution of the perturbations.
	 In this case, the perturbed energy–momentum tensor on the right‑hand side of the linearized Einstein equations vanishes, \(\delta T_{\mu\nu}=0\), and the equations take the same form as in vacuum. Following the common phenomenological approach widely employed in the study of non‑vacuum or modified‑gravity black holes~\cite{Gogoi:2024epx,Liu:2024xcd,Liang:2024geh}, all the correction effects are entirely encoded in the background metric function \(f(r)\), while perturbations are treated as propagating on this curved spacetime background. Consequently, the effective potential for the axial gravitational perturbation is taken as the direct generalization of the vacuum Regge–Wheeler potential:
	\begin{equation}
		V_{g}(r)=f(r)\left[\frac{l(l+1)}{r^{2}}-\frac{3}{r}\frac{df(r)}{dr}\right]
		\label{eq:20}
	\end{equation}
	~~where the angular quantum number satisfies \(l\ge 2\). This form automatically reduces to the standard result in the Schwarzschild limit. In the non‑vacuum background considered here, the effects of Lorentz violation, electric charge, and dark matter on the spacetime geometry are all encoded in the metric function \(f(r)\) and its derivative, and they manifest themselves through the effective potential.
	
		\begin{figure*}[htbp]
		\centering
		\renewcommand{\thesubfigure}{\roman{subfigure}}  
		\subcaptionbox{$\tau=0.1$, $\lambda/M=0.2$}{
			\includegraphics[width=0.3\textwidth]{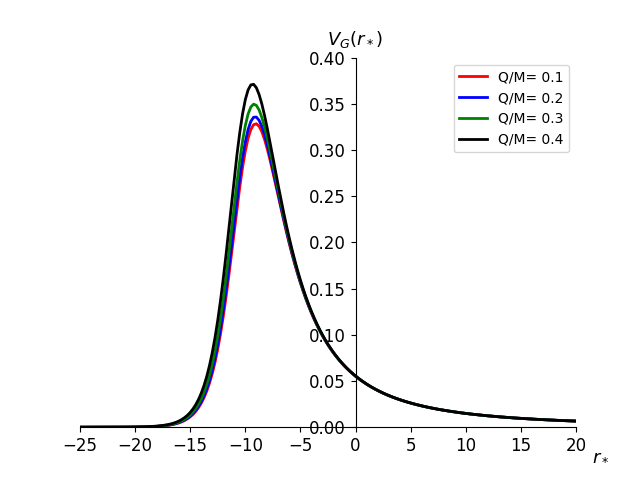}
		}
		\hfill
		\subcaptionbox{$\tau=0.1$, $Q/M=0.5$}{
			\includegraphics[width=0.3\textwidth]{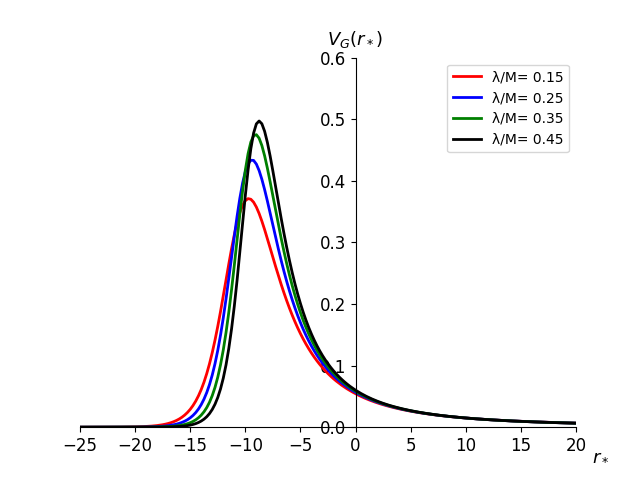}
		}
		\hfill
		\subcaptionbox{$\lambda/M=0.2$, $Q/M=0.5$}{
			\includegraphics[width=0.3\textwidth]{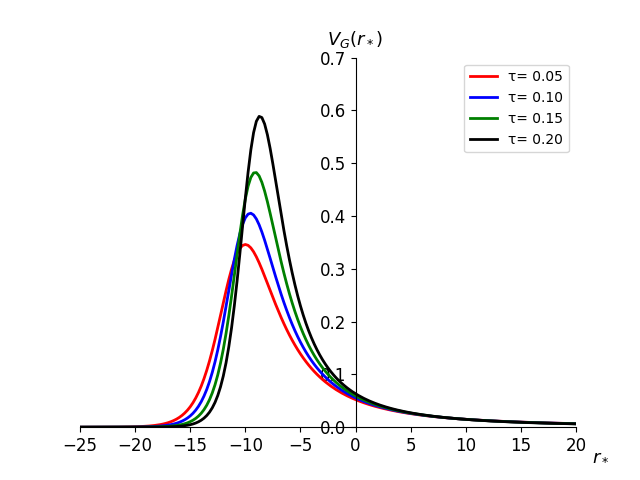}
		}
		\caption{Variation of the effective potential $V_G(r_*)$ with the tortoise coordinate $r_*$ for a charged KR black hole immersed in PFDM under a gravitational field perturbation ($M=1$, $l=2$).}
		\label{fig:5}
	\end{figure*}
	
	Summarizing the above derivations, the effective potentials for the three types of perturbations can be uniformly expressed as
	\begin{equation}
		V_{s}(r) = f(r)\left[\frac{l(l+1)}{r^{2}}+\frac{1-s^{2}}{r}\frac{df(r)}{dr}\right]
		\label{eq:21}
	\end{equation}
	~~where $s=0,1,2$ correspond to the scalar field, electromagnetic field, and axial gravitational perturbation, respectively. This unified form encompasses the classical results for massless fields in static spherically symmetric spacetimes. 
	
	~~In the subsequent calculations, based on the Schr\"{o}dinger-like equation and its boundary conditions, we will employ the sixth-order WKB method and the time-domain evolution method to solve the QNM frequencies, and systematically investigate the influence of the Lorentz-violating parameter $\tau$, the charge $Q/M$, and the dark matter parameter $\lambda/M$ on the QNM spectrum.
	
		\begin{figure*}[htbp]
		\centering
		\renewcommand{\thesubfigure}{\roman{subfigure}}  
		\subcaptionbox{$\tau=0.1$, $\lambda/M=0.2$}{
			\includegraphics[width=0.3\textwidth]{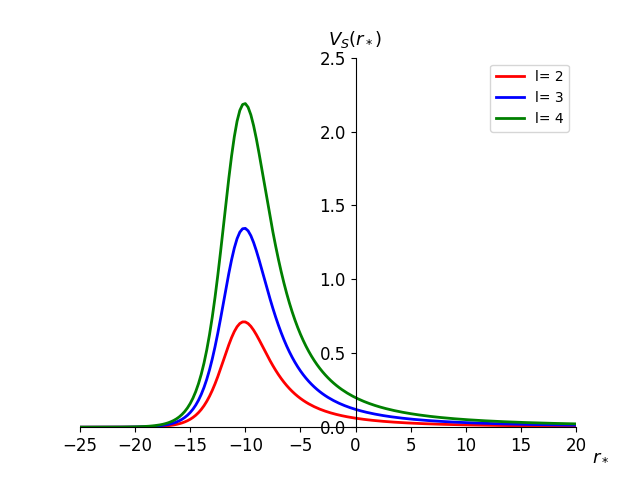}
		}
		\hfill
		\subcaptionbox{$\tau=0.1$, $Q/M=0.5$}{
			\includegraphics[width=0.3\textwidth]{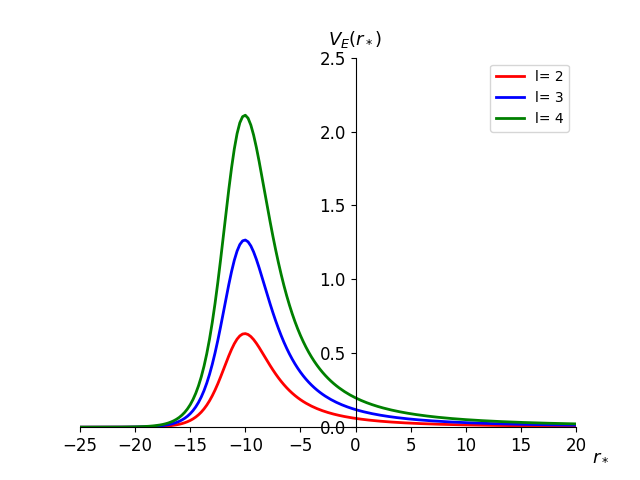}
		}
		\hfill
		\subcaptionbox{$\lambda/M=0.2$, $Q/M=0.5$}{
			\includegraphics[width=0.3\textwidth]{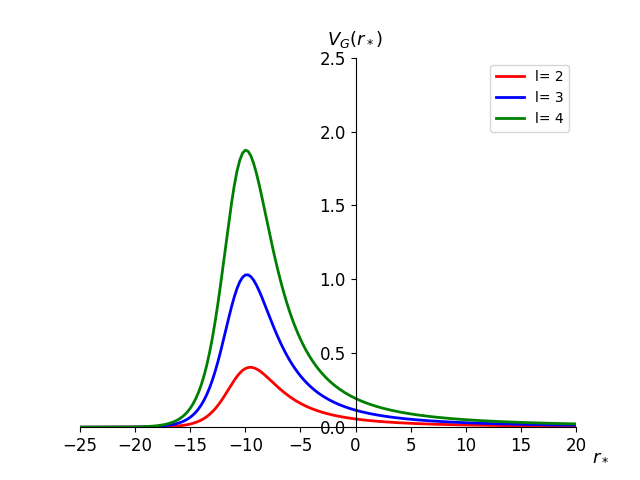}
		}
		\caption{Effective potential barrier for different types of perturbations with fixed parameters $M=1$, $\tau=0.1$, $\lambda=0.2$, $Q=0.5$, illustrating the influence of $l$ on the effective potential barrier.}
		\label{fig:6}
	\end{figure*}
	~~In all the parameter scans displayed in Figs.~\ref{fig:3}, \ref{fig:4}, and \ref{fig:5} the peak value of the effective potential barrier increases monotonically with each parameter, indicating an enhanced confining ability of the barrier. The shift of the peak position depends simultaneously on the type of parameter and on the spin of the perturbing field. Briefly, increasing the charge parameter \(Q/M\) causes a slight shift of the peak toward the horizon (leftward movement) in the tortoise coordinate, whereas increasing the dark matter parameter \(\lambda/M\) or the Lorentz-violating parameter \(\tau\) leads to a pronounced shift of the peak away from the horizon (rightward movement) in the tortoise coordinate. This difference originates from the fact that the charge term mainly produces corrections in the near-horizon region, while the dark matter term and the Lorentz-violating term generate an overall uplifting effect on the entire exterior spacetime. Overall, the charge parameter significantly regulates only the height of the potential barrier for the three types of perturbations and has very little influence on the peak position, whereas the dark matter and Lorentz-violating parameters regulate both the height and the peak position.
	
	~~ Fig.\ref{fig:6} shows that the effective potentials for the three types of perturbations all exhibit a typical single-peak structure and tend to zero both at the horizon and at infinity, satisfying the asymptotic boundary conditions for QNMs. As \(l\) increases from 2 to 4, the peak height of the potential barrier rises monotonically. This trend is dominated by the centrifugal barrier term \(f(r)l(l+1)/r^{2}\): the larger the angular quantum number \(l\), the stronger the centrifugal contribution, and the peak height increases significantly with \(l(l+1)\). We note that, under the same parameters, the effective potential barriers for the scalar and electromagnetic perturbations are similar in shape, but the scalar barrier is slightly higher. This is because the scalar effective potential contains an additional curvature coupling term \(f(r)f'(r)/r\). As the angular quantum number \(l\) increases, the centrifugal barrier term gradually dominates, and the difference between the two types of perturbations becomes smaller.
	
	\subsection{Calculation methods and numerical results of quasinormal modes}
	\label{sec:3.4}
	The damped oscillations of a perturbed black hole are described by quasinormal modes (QNMs), whose complex frequencies \(\omega = \omega_{\mathrm{R}} - i\omega_{\mathrm{I}}\) depend only on the geometric properties of the background spacetime and are independent of the initial perturbation. They therefore serve as a central tool for probing black hole parameters and testing gravitational theories. In this section, we combine two independent methods to compute the QNMS of a charged KR black hole immersed in PFDM under scalar, electromagnetic, and axial gravitational perturbations: one is the semi-analytic sixth-order WKB method, and the other is the time-domain evolution method for extracting eigenvalues. The mutual verification of the two methods ensures the reliability of the results.
	
	~~ The wave equations for the three types of perturbations have been derived in the previous sections. By separating the time variable, the three wave equations can be further reduced to the following time-independent radial equation:
	\begin{equation}
		\frac{d^{2}\psi(r_{*})}{dr_{*}^{2}} + \bigl[\omega^{2} - V(r_{*})\bigr]\psi(r_{*}) = 0
		\label{eq:22}
	\end{equation}
	~~The QNM frequencies \(\omega\) are discrete complex frequencies determined by specific boundary conditions, namely that the radial wave function is purely ingoing at the horizon (\(r_{*} \to -\infty\)) and purely outgoing at spatial infinity (\(r_{*} \to +\infty\)). 
	It should be pointed out that the background spacetime considered here is not asymptotically flat (\(f(\infty) = 1/(1-\tau) \neq 1\)). Nevertheless, since the effective potential decays to zero at infinity in terms of the tortoise coordinate, the form of the QNM boundary conditions remains the same as in the asymptotically flat case, i.e., \(\psi \sim e^{-i\omega r_{*}}\ (r_{*} \to -\infty)\) and \(\psi \sim e^{i\omega r_{*}}\ (r_{*} \to +\infty)\). The effect of non-asymptotic flatness has been fully absorbed into the mapping between the tortoise coordinate and the radial coordinate.
	
	~~When the effective potential has a single-peak, the WKB semi-classical approximation can be employed. Expanding around the peak of the potential barrier and matching the asymptotic solutions on both sides yields the following equation~\cite{Konoplya:2003dd,Konoplya:2003ii,Iyer:1986nq}:
	\begin{equation}
		\frac{i(\omega^{2}-V_{0})}{\sqrt{-2V_{0}''}} - \sum_{k=2}^{6}\Lambda_{k} = n + \frac{1}{2}
		\label{eq:23}
	\end{equation}
	~~where \(V_{0}\) and \(V_{0}''\) denote the value of the effective potential at its peak and its second derivative with respect to the tortoise coordinate \(r_{*}\), respectively, \(\Lambda_{k}\) is the \(k\)-th order WKB correction term, and \(n\) is the overtone number. The applicability condition of the WKB method is \(l > n\). For the fundamental mode \(n=0\), the method yields good accuracy for \(l \ge 1\)~\cite{Konoplya:2019hlu}.

	~~ In addition to employing the sixth-order WKB method to calculate the QNM frequencies, we can also use the time-domain evolution method to directly evolve the original wave equation numerically.
	
	In terms of the tortoise coordinate \(r_{*}\), the radial part of the perturbation field \(\psi(r_{*},t)\) satisfies the wave equation
	\begin{equation}
		\frac{\partial^{2}\psi}{\partial t^{2}} - \frac{\partial^{2}\psi}{\partial r_{*}^{2}} + V(r_{*})\psi = 0
		\label{eq:24}
	\end{equation}
	~~Upon introducing the light-cone coordinates \(u = t - r_{*}\) and \(v = t + r_{*}\), the wave equation~\eqref{eq:24} can be recast as
	\begin{equation}
		4\frac{\partial^{2}\psi}{\partial u\,\partial v} =  - V(r_{*})\psi
		\label{eq:25}
	\end{equation}
	~~Taking a uniform grid on the \((u,v)\) plane with spacing \(\Delta u = \Delta v = h\), the grid points are denoted as \(\psi_{i,j} = \psi(u_{i}, v_{j})\), where
	\(u_{i} = u_{0} + ih\) and \(v_{j} = v_{0} + jh\)
	
	~~To solve Eq.~\eqref{eq:25}, it can be discretized into the following form~\cite{Gundlach:1993tp}:
	\begin{equation}
		\begin{split}
			\psi_{i+1,j+1} &= \psi_{i+1,j} + \psi_{i,j+1} - \psi_{i,j} \\
			&\quad - \frac{h^{2}}{8} V_{i,j}\bigl(\psi_{i+1,j} + \psi_{i,j+1}\bigr) + \mathcal{O}(h^{4})
		\end{split}
		\label{eq:26}
	\end{equation}
	~~where \(V_{i,j} = V\bigl(r_{*}(u_{i},v_{j})\bigr)\) and \(\mathcal{O}(h^{4})\) denotes the higher-order infinitesimal term. 
	
	~~The initial condition is introduced as a Gaussian pulse~\cite{Moderski:2001gt}:
	\begin{equation}
		\begin{cases}
			\psi(u = u_{0}, v) = A \exp\!\left(-\dfrac{(v - v_{c})^{2}}{2\sigma^{2}}\right) \\[8pt]
			\psi(u, v = v_{0}) = 0
		\end{cases}
		\label{eq:27}
	\end{equation}
	~~where \(A\) is the amplitude, \(\sigma\) is the width of the wave packet, and \(v_{c}\) specifies the center of the wave packet. This setup imposes a Gaussian-distributed initial perturbation along the \(v\)-direction at \(u = u_{0}\). In terms of the tortoise coordinate, this corresponds to a wave packet propagating from a distant region toward the horizon, while setting the field to zero at \(v = v_{0}\) ensures that there is no initial outgoing wave from the direction of the horizon.
	
	~~ During the time evolution, the time-domain waveform exhibits three characteristic stages: the initial transient whose shape depends on the form of the initial perturbation, the ringdown stage dominated entirely by the black hole geometry, and the late-time power-law tail. To accurately extract the QNM frequencies, we retain only the data from the ringdown stage, excluding the initial transient and the late-time power-law tail, and apply the Prony method~\cite{Konoplya:2011qq,Chowdhury:2020rfj,Berti:2007dg} to perform a complex exponential fitting of the signal, from which the frequencies and damping rates of the QNMs are obtained:
	\begin{equation}
		\psi(t) \simeq \sum_{j=1}^{k} C_{j} e^{-i\omega_{j} t}
		\label{eq:28}
	\end{equation}
	~~where \(C_{j}\) denotes the complex amplitude and \(k\) is the number of fitting modes.
	
\begin{figure*}[htbp]
		\centering
		\includegraphics[width=0.45\textwidth]{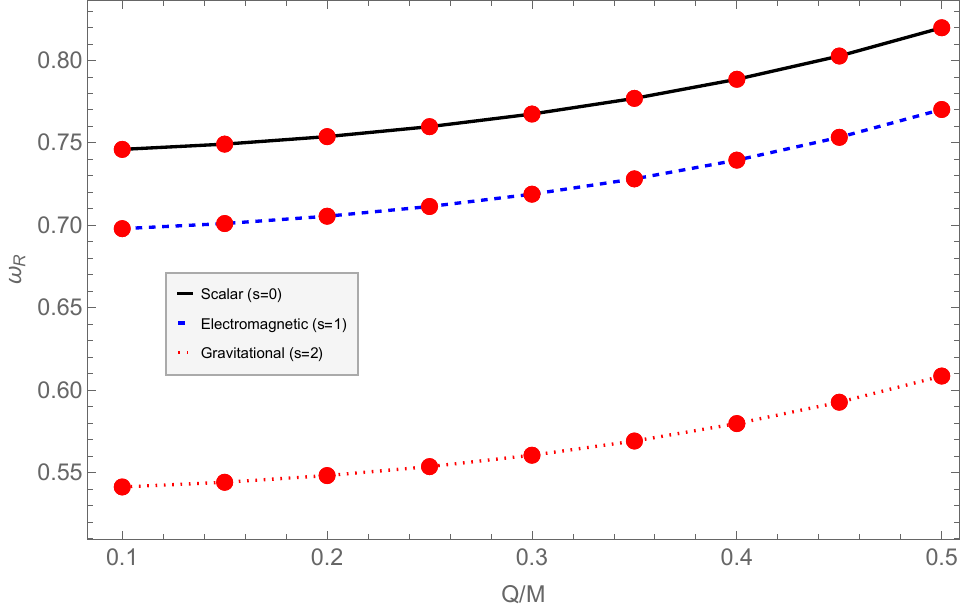}\hfill
		\includegraphics[width=0.45\textwidth]{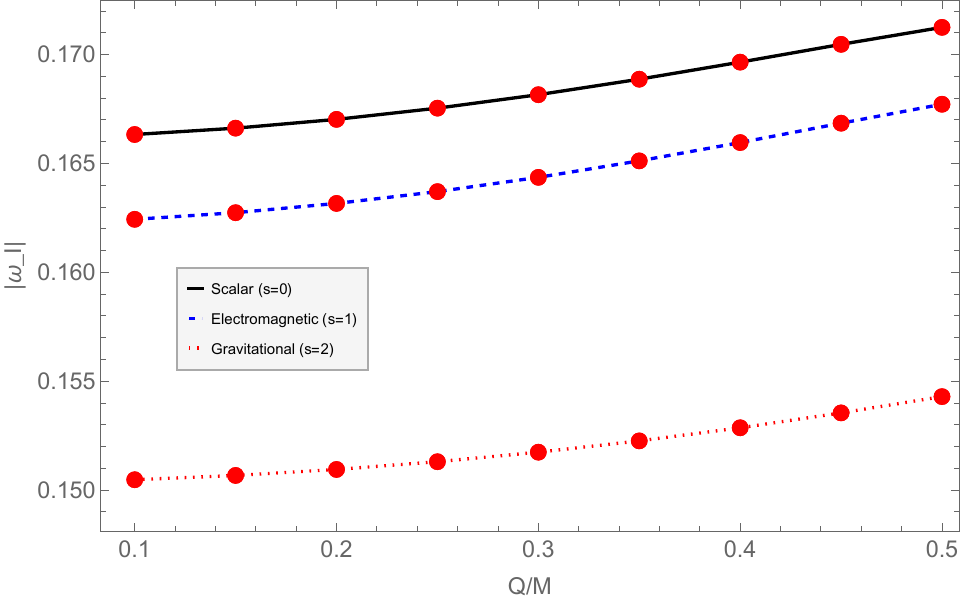}
		\caption{Variations of the real‑part and absolute value of the imaginary‑part of QNM frequencies against different values of \(Q/M\) for three types of perturbation fields, with fixed parameters \(l=2\), \(M=1\), \(\tau=0.1\) and \(\lambda/M=0.2\).}
		\label{fig:11}
	\end{figure*}

\begin{figure*}[htbp]
		\centering
		\includegraphics[width=0.45\textwidth]{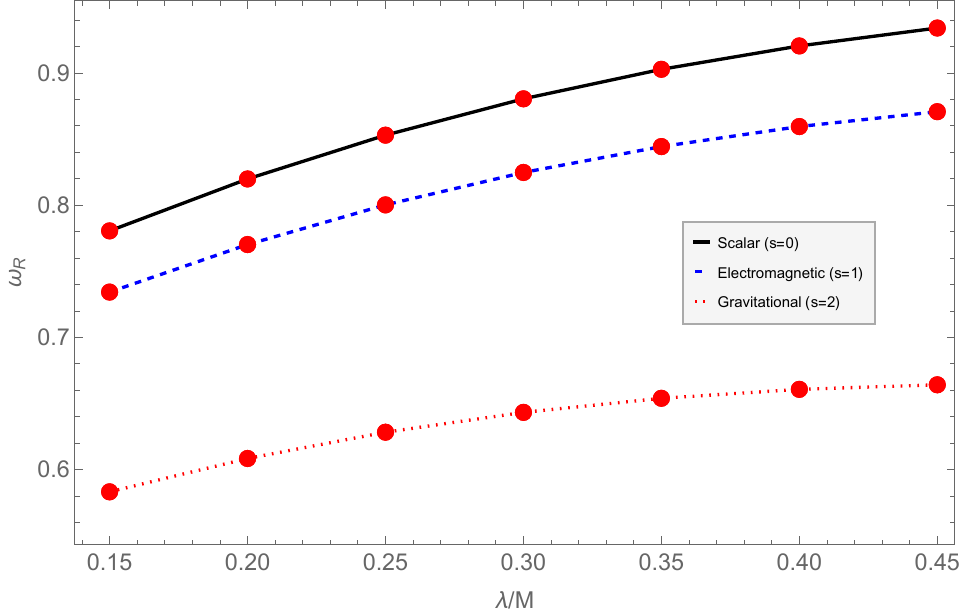}\hfill
		\includegraphics[width=0.45\textwidth]{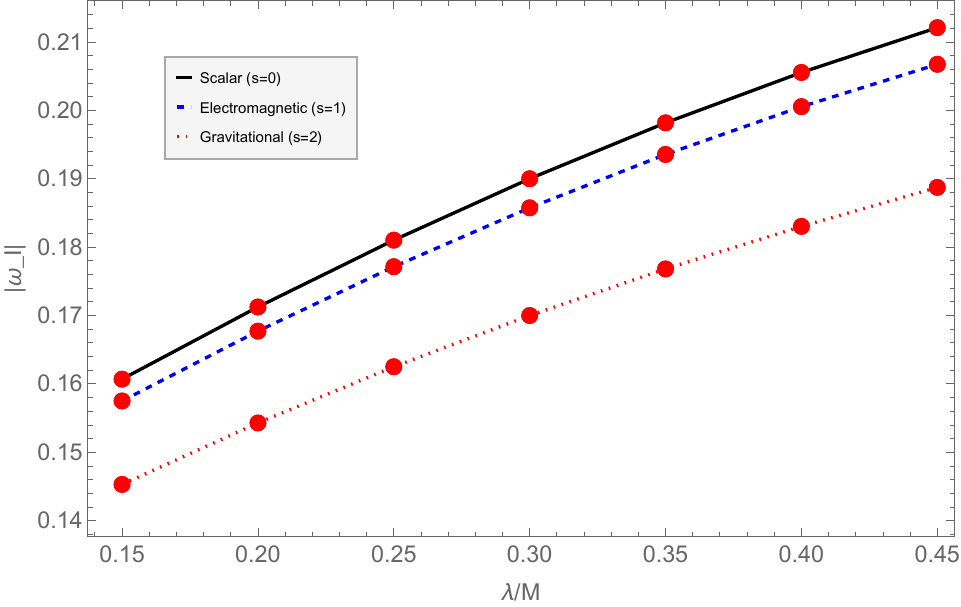}
		\caption{Variations of the real‑part and absolute value of the imaginary‑part of QNM frequencies against different values of \(\lambda/M\) for three types of perturbation fields, with fixed parameters \(l=2\), \(M=1\), \(\tau=0.1\) and \(Q/M=0.5\).}
		\label{fig:12}
	\end{figure*}

\begin{figure*}[tp]
	\centering
	\includegraphics[width=0.45\textwidth]{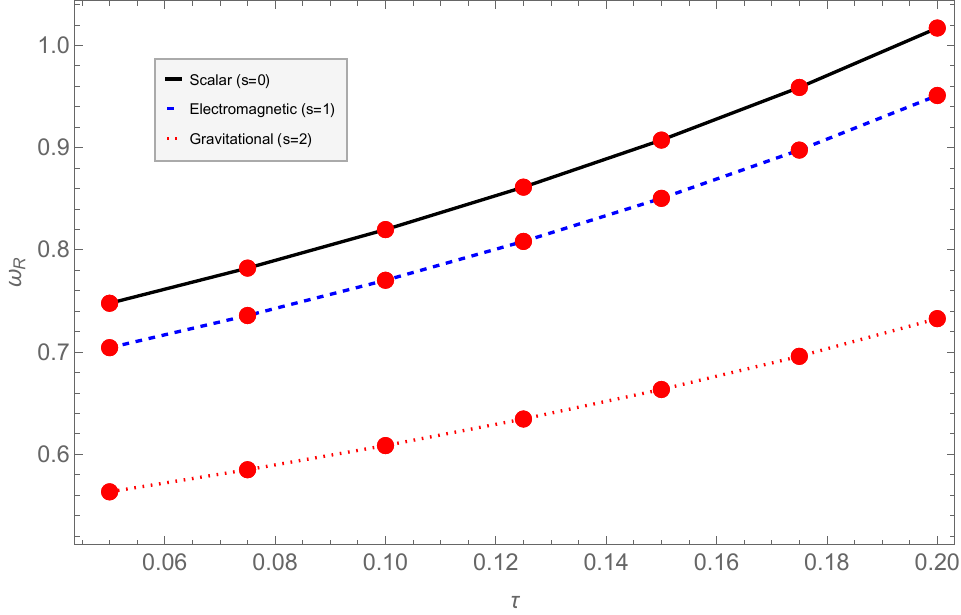}\hfill
	\includegraphics[width=0.45\textwidth]{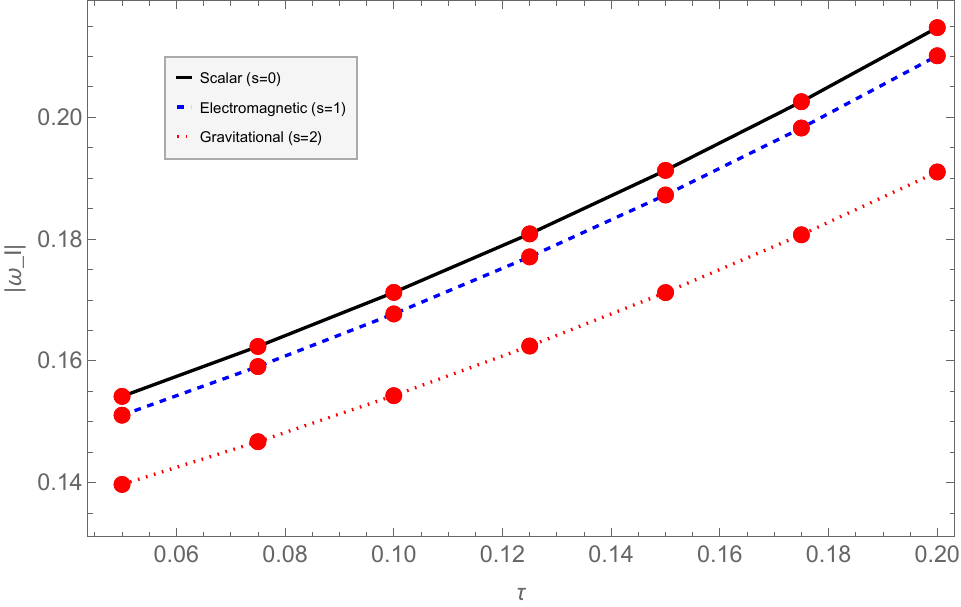}
	\caption{Variations of the real‑part and absolute value of the imaginary‑part of QNM frequencies against different values of \(\tau\) for three types of perturbation fields, with fixed parameters \(l=2\), \(M=1\), \(\lambda/M=0.2\) and \(Q/M=0.5\).}
	\label{fig:13}
\end{figure*}

\begin{figure*}[htbp]
	\centering
	\includegraphics[width=0.45\textwidth]{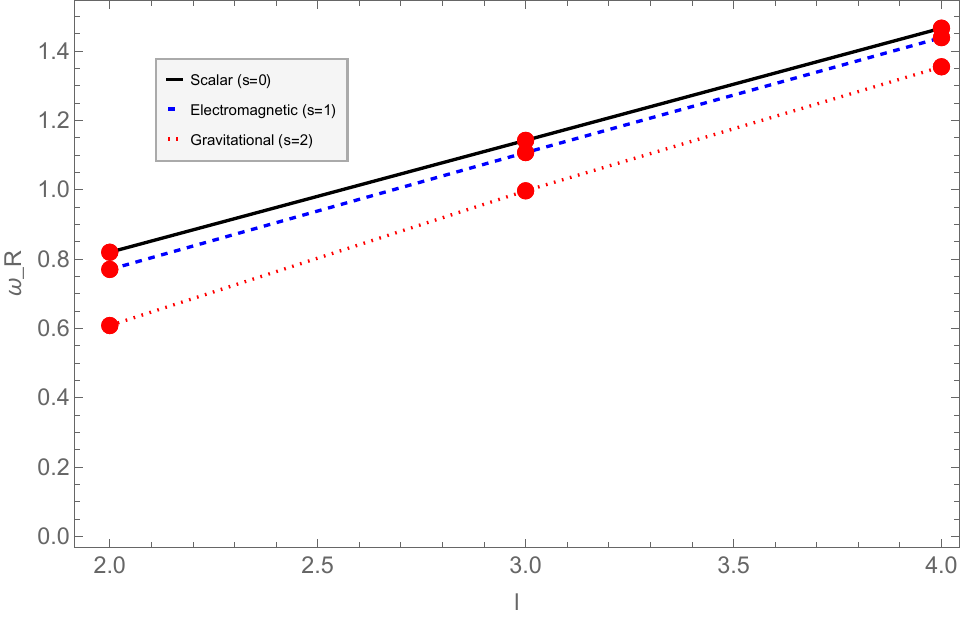}\hfill
	\includegraphics[width=0.45\textwidth]{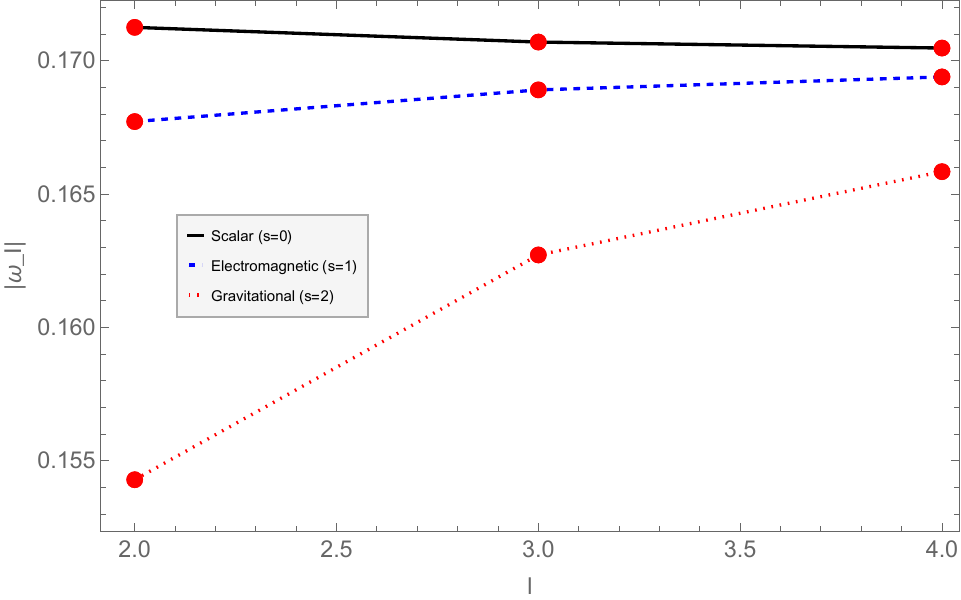}
	\caption{Variations of the real‑part and absolute value of the imaginary‑part of QNM frequencies against different values of \(l\) for three types of perturbation fields, with fixed parameters \(M=1\), \(\lambda/M=0.2\), \(\tau=0.1\) and \(Q/M=0.5\).}
	\label{fig:14}
\end{figure*}

		Tables~\ref{tab:1}--\ref{tab:4} and Figs.~\ref{fig:11}--\ref{fig:14} systematically present the QNM frequencies of the charged KR--PFDM black hole under scalar, electromagnetic, and axial gravitational perturbations. The results obtained from two independent methods, the WKB approximation and the Prony method, are in excellent agreement with each other, which fully confirms the reliability of the numerical calculations. Under the same parameters, the oscillation frequencies \(\omega_R\) and the damping rates \(|\omega_I|\) of the three types of perturbations always satisfy the ordering scalar field \(>\) electromagnetic field \(>\) axial gravitational field, consistent with the ordering of the effective potential heights shown in Figs.~\ref{fig:3}--\ref{fig:6}.
	
\begin{figure*}[htbp]
		\centering
		\includegraphics[width=0.3\textwidth]{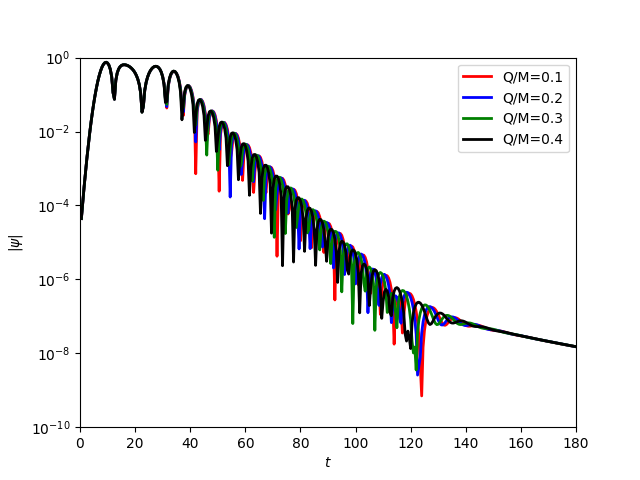}\hfill
		\includegraphics[width=0.3\textwidth]{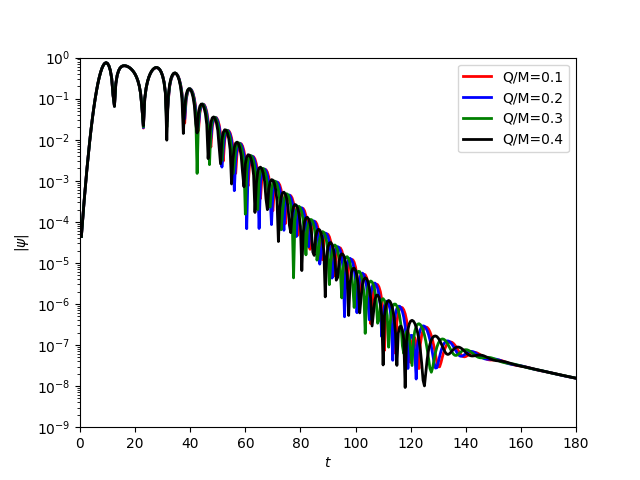}\hfill
		\includegraphics[width=0.3\textwidth]{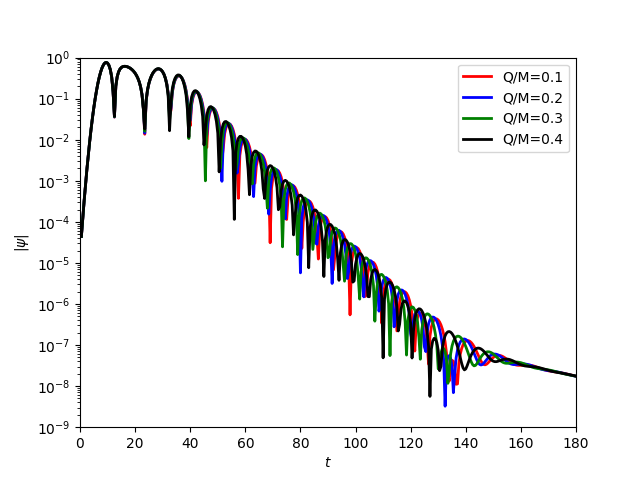}
		\caption{Time-domain evolution for scalar, electromagnetic, and axial gravitational perturbations (left to right) under different \(Q/M\), with \(l=2\), \(M=1\), \(\tau=0.1\), and \(\lambda/M=0.2\).}
		\label{fig:7}
	\end{figure*}

\begin{figure*}[htbp]
		\centering
		\includegraphics[width=0.3\textwidth]{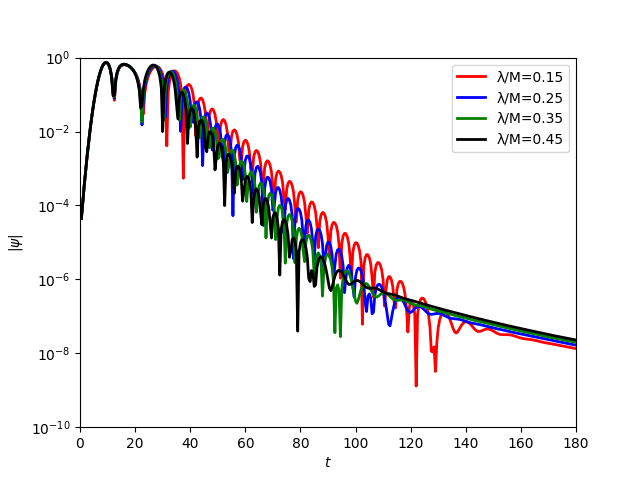}\hfill
		\includegraphics[width=0.3\textwidth]{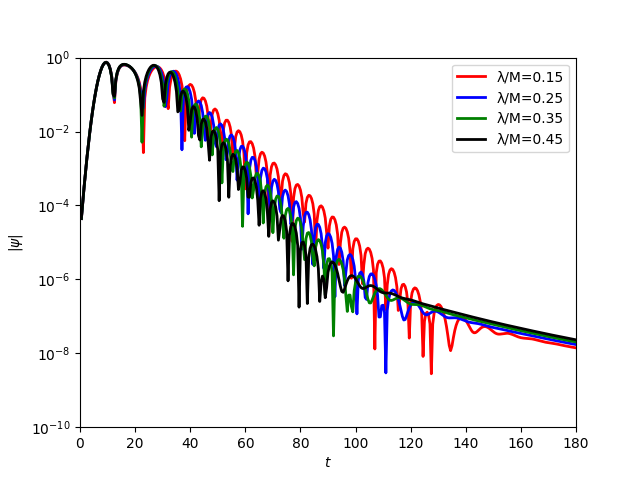}\hfill
		\includegraphics[width=0.3\textwidth]{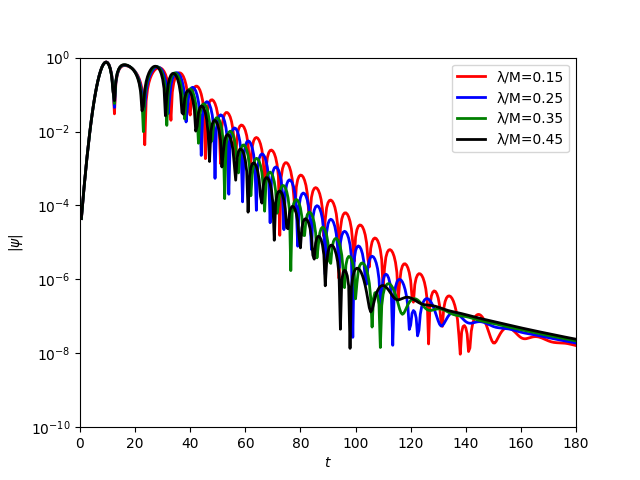}
		\caption{Time-domain evolution for scalar, electromagnetic, and axial gravitational perturbations (left to right) under different \(\lambda/M\), with \(l=2\), \(M=1\), \(\tau=0.1\), and \(Q/M=0.5\).}
		\label{fig:8}
	\end{figure*}

\begin{figure*}[htbp]
		\centering
		\includegraphics[width=0.3\textwidth]{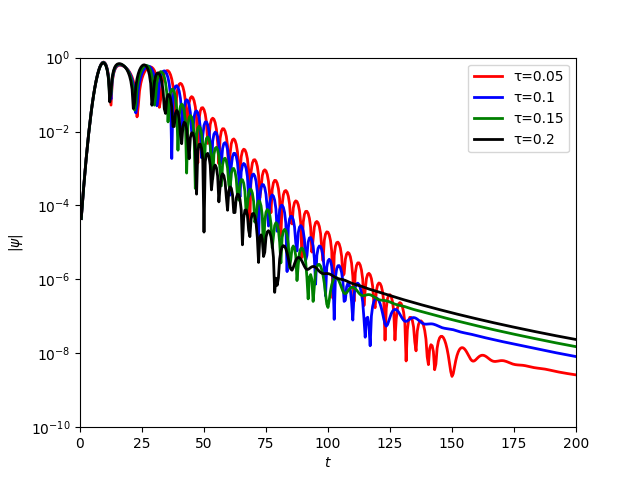}\hfill
		\includegraphics[width=0.3\textwidth]{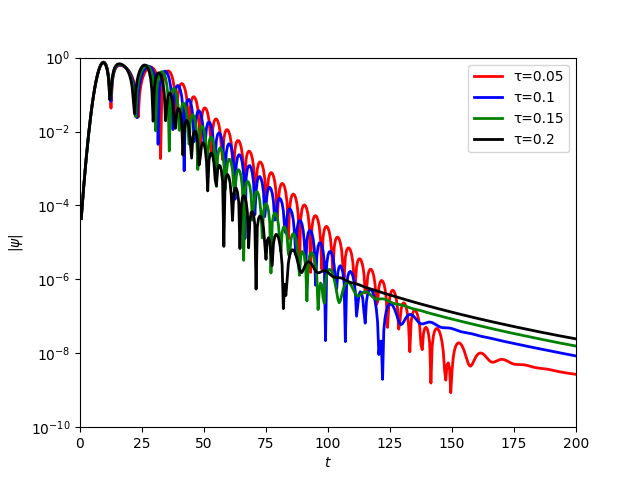}\hfill
		\includegraphics[width=0.3\textwidth]{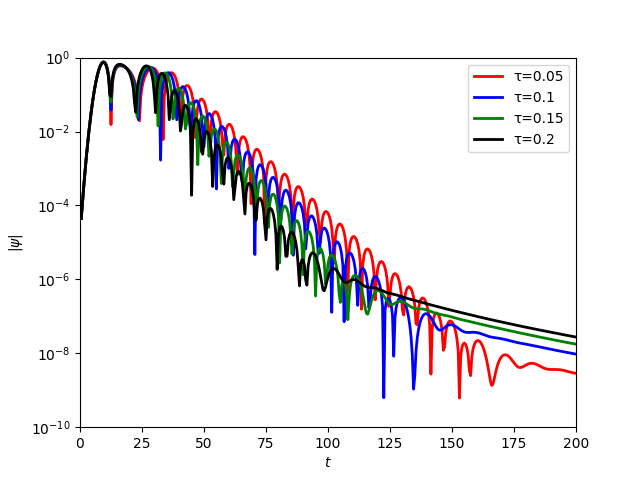}
		\caption{Time-domain evolution for scalar, electromagnetic, and axial gravitational perturbations (left to right) under different \(\tau\), with \(l=2\), \(M=1\), \(\lambda/M=0.2\), and \(Q/M=0.5\).}
		\label{fig:9}
	\end{figure*}

\begin{figure*}[htbp]
		\centering
		\includegraphics[width=0.3\textwidth]{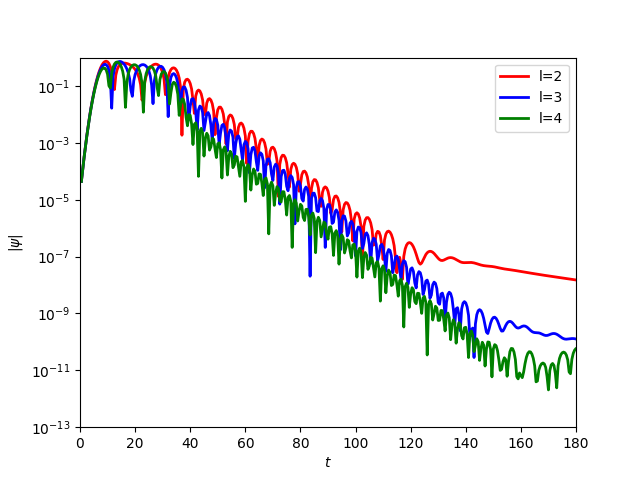}\hfill
		\includegraphics[width=0.3\textwidth]{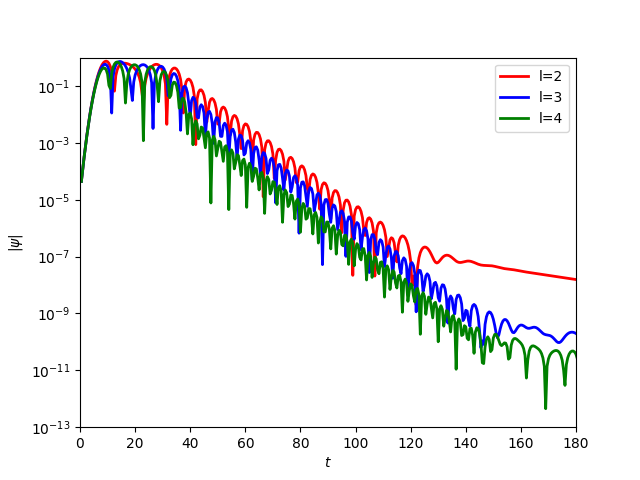}\hfill
		\includegraphics[width=0.3\textwidth]{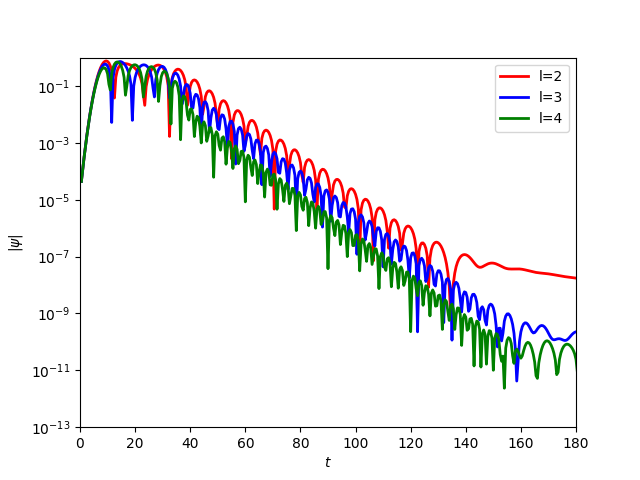}
		\caption{Time-domain evolution for scalar, electromagnetic, and axial gravitational perturbations (left to right) under different \(l\), with \(M=1\), \(\lambda/M=0.2\), \(Q/M=0.5\), and \(\tau=0.1\).}
		\label{fig:10}
	\end{figure*}

	By combining the information from Figs.~\ref{fig:7}--\ref{fig:14} and Tables~\ref{tab:1}--\ref{tab:4}, the following features can be observed. The charge parameter \(Q/M\) has a relatively weak effect on the QNM frequencies. As \(Q/M\) increases, both \(\omega_R\) and \(|\omega_I|\) of the three types of perturbations rise only slightly, and the corresponding changes in the ringdown frequency and damping rate in the time-domain waveforms are not obvious. This is consistent with the earlier observation that the effective potential barrier is only mildly elevated by increasing \(Q/M\). It originates from the fact that the modification of the metric function by the charge term is mainly concentrated in the near-horizon region and has limited influence on the spacetime curvature around the peak of the potential barrier. The PFDM parameter \(\lambda/M\) exerts a more pronounced influence on the QNM frequencies. As \(\lambda/M\) increases, both \(\omega_R\) and \(|\omega_I|\) of the three types of perturbations increase significantly, and the ringdown frequency and damping rate in the time-domain waveforms become distinctly faster, in agreement with the evolutionary trend of the effective potential barrier with \(\lambda/M\) described previously. The Lorentz-violating parameter \(\tau\) has the most prominent effect on the QNM frequencies. As \(\tau\) increases, both \(\omega_R\) and \(|\omega_I|\) of the three perturbations rise substantially, and the variations of the ringdown frequency and damping rate in the time-domain waveforms are more drastic than those in the \(\lambda\) case. The physical origin lies in the fact that \(\tau\) modifies the constant term and the charge term in the metric function via the factors \((1-\tau)^{-1}\) and \((1-\tau)^{-2}\), leading to an overall enhancement of the spacetime curvature and a significant elevation of the effective potential barrier.
	
	The influence of the angular quantum number \(l\) on the QNM frequencies exhibits characteristics different from those of the above parameters. As \(l\) increases, \(\omega_R\) of the three types of perturbations rises markedly, while the change in \(|\omega_I|\) is very small. In the time-domain waveforms, the ringdown frequency of high-\(l\) modes is clearly higher, but the damping rate changes little. This trend is consistent with the behavior of the centrifugal barrier term \(f(r)l(l+1)/r^2\) with increasing \(l\) (see Fig.~\ref{fig:6}).
	
	Overall, the QNM spectrum of the charged KR--PFDM black hole is more sensitive to variations in \(\lambda/M\) and \(\tau\), while it is less affected by \(Q/M\). The above results provide a theoretical reference for distinguishing the effects of Lorentz violation and dark matter through gravitational wave observations.

	
\clearpage   

\begin{table*}[htbp]
	\centering
	\footnotesize
	\setlength{\tabcolsep}{4pt}
	\begin{tabular}{c c c c c c c}
		\hline
		& \multicolumn{2}{c}{Scalar} & \multicolumn{2}{c}{Electromagnetic} & \multicolumn{2}{c}{Axial gravitational} \\
		\hline
		& WKB & Prony & WKB & Prony & WKB & Prony \\
		\hline
		\(Q/M\) & \multicolumn{6}{c}{\(\tau=0.1,\ \lambda/M=0.2,\ n=0,\ l=2,\ M=1\)} \\
		\hline
		0.1 & \(0.746037 - 0.166327\,\mathrm{i}\) & \(0.746986 - 0.165381\,\mathrm{i}\)
		& \(0.697936 - 0.162432\,\mathrm{i}\) & \(0.698853 - 0.161515\,\mathrm{i}\)
		& \(0.541206 - 0.150474\,\mathrm{i}\) & \(0.543109 - 0.149554\,\mathrm{i}\) \\
		0.2 & \(0.753788 - 0.167021\,\mathrm{i}\) & \(0.754768 - 0.166081\,\mathrm{i}\)
		& \(0.705487 - 0.163161\,\mathrm{i}\) & \(0.706410 - 0.162219\,\mathrm{i}\)
		& \(0.548130 - 0.150946\,\mathrm{i}\) & \(0.549855 - 0.150174\,\mathrm{i}\) \\
		0.3 & \(0.767490 - 0.168154\,\mathrm{i}\) & \(0.768534 - 0.167211\,\mathrm{i}\)
		& \(0.718860 - 0.164358\,\mathrm{i}\) & \(0.719786 - 0.163380\,\mathrm{i}\)
		& \(0.560466 - 0.151738\,\mathrm{i}\) & \(0.561885 - 0.151181\,\mathrm{i}\) \\
		0.4 & \(0.788582 - 0.169648\,\mathrm{i}\) & \(0.789725 - 0.168685\,\mathrm{i}\)
		& \(0.739508 - 0.165956\,\mathrm{i}\) & \(0.740425 - 0.164962\,\mathrm{i}\)
		& \(0.579672 - 0.152860\,\mathrm{i}\) & \(0.580689 - 0.152503\,\mathrm{i}\) \\
		\hline
	\end{tabular}
	\caption{QNM frequencies for different values of \(Q/M\) obtained using the sixth‑order WKB approximation and the Prony method.}
	\label{tab:1}
\end{table*}
	
\begin{table*}[htbp]
	\centering
	\footnotesize
	\setlength{\tabcolsep}{4pt}
	\begin{tabular}{c c c c c c c}
		\hline
		& \multicolumn{2}{c}{Scalar} & \multicolumn{2}{c}{Electromagnetic} & \multicolumn{2}{c}{Axial gravitational} \\
		\hline
		& WKB & Prony & WKB & Prony & WKB & Prony \\
		\hline
		\(\lambda/M\) & \multicolumn{6}{c}{\(\tau=0.1,\ Q/M=0.5,\ n=0,\ l=2,\ M=1\)} \\
		\hline
		0.15 & \(0.780570 - 0.160692\,\mathrm{i}\) & \(0.781988 - 0.159733\,\mathrm{i}\)
		& \(0.734308 - 0.157504\,\mathrm{i}\) & \(0.735517 - 0.156638\,\mathrm{i}\)
		& \(0.583291 - 0.145297\,\mathrm{i}\) & \(0.583863 - 0.144613\,\mathrm{i}\) \\
		0.25 & \(0.852945 - 0.181017\,\mathrm{i}\) & \(0.854507 - 0.179907\,\mathrm{i}\)
		& \(0.800222 - 0.177131\,\mathrm{i}\) & \(0.799205 - 0.176191\,\mathrm{i}\)
		& \(0.628356 - 0.162509\,\mathrm{i}\) & \(0.629060 - 0.162135\,\mathrm{i}\) \\
		0.35 & \(0.902731 - 0.198171\,\mathrm{i}\) & \(0.904085 - 0.198382\,\mathrm{i}\)
		& \(0.844319 - 0.193548\,\mathrm{i}\) & \(0.844843 - 0.193107\,\mathrm{i}\)
		& \(0.653999 - 0.176808\,\mathrm{i}\) & \(0.655261 - 0.176386\,\mathrm{i}\) \\
		0.45 & \(0.933939 - 0.212092\,\mathrm{i}\) & \(0.936506 - 0.210228\,\mathrm{i}\)
		& \(0.870688 - 0.206726\,\mathrm{i}\) & \(0.870787 - 0.204779\,\mathrm{i}\)
		& \(0.664210 - 0.188730\,\mathrm{i}\) & \(0.666553 - 0.187818\,\mathrm{i}\) \\
		\hline
	\end{tabular}
	\caption{QNM frequencies for different values of $\lambda/M$ obtained using the sixth‑order WKB approximation and the Prony method.}
	\label{tab:2}
\end{table*}
	
	\begin{table*}[htbp]
		\centering
		\footnotesize
		\setlength{\tabcolsep}{4pt}
		\begin{tabular}{c c c c c c c}
			\hline
			& \multicolumn{2}{c}{Scalar} & \multicolumn{2}{c}{Electromagnetic} & \multicolumn{2}{c}{Axial gravitational} \\
			\hline
			& WKB & Prony & WKB & Prony & WKB & Prony \\
			\hline
			\(\tau\) & \multicolumn{6}{c}{\(\lambda/M=0.2,\ Q/M=0.5,\ n=0,\ l=2,\ M=1\)} \\
			\hline
			0.05 & \(0.747746 - 0.154167\,\mathrm{i}\) & \(0.748990 - 0.153317\,\mathrm{i}\)
			& \(0.704283 - 0.151080\,\mathrm{i}\) & \(0.704585 - 0.150143\,\mathrm{i}\)
			& \(0.563255 - 0.139706\,\mathrm{i}\) & \(0.563983 - 0.138640\,\mathrm{i}\) \\
			0.10 & \(0.819834 - 0.171246\,\mathrm{i}\) & \(0.821518 - 0.170102\,\mathrm{i}\)
			& \(0.770250 - 0.167715\,\mathrm{i}\) & \(0.771213 - 0.166773\,\mathrm{i}\)
			& \(0.608502 - 0.154291\,\mathrm{i}\) & \(0.609110 - 0.153905\,\mathrm{i}\) \\
			0.15 & \(0.907406 - 0.191290\,\mathrm{i}\) & \(0.909163 - 0.190539\,\mathrm{i}\)
			& \(0.850414 - 0.187246\,\mathrm{i}\) & \(0.852178 - 0.185976\,\mathrm{i}\)
			& \(0.663375 - 0.171221\,\mathrm{i}\) & \(0.663746 - 0.170637\,\mathrm{i}\) \\
			0.20 & \(1.017040 - 0.214739\,\mathrm{i}\) & \(1.020240 - 0.212578\,\mathrm{i}\)
			& \(0.951039 - 0.210117\,\mathrm{i}\) & \(0.948455 - 0.210903\,\mathrm{i}\)
			& \(0.732713 - 0.191031\,\mathrm{i}\) & \(0.733312 - 0.189737\,\mathrm{i}\) \\
			\hline
		\end{tabular}
		\caption{QNM frequencies for different values of \(\tau\) obtained using the sixth‑order WKB approximation and the Prony method.}
		\label{tab:3}
	\end{table*}
	
	\begin{table*}[htbp]
		\centering
		\footnotesize
		\setlength{\tabcolsep}{4pt}
		\begin{tabular}{c c c c c c c}
			\hline
			& \multicolumn{2}{c}{Scalar} & \multicolumn{2}{c}{Electromagnetic} & \multicolumn{2}{c}{Axial gravitational} \\
			\hline
			& WKB & Prony & WKB & Prony & WKB & Prony \\
			\hline
			\(l\) & \multicolumn{6}{c}{\(Q/M=0.5,\ \lambda/M=0.2,\ \tau=0.1,\ n=0,\ M=1\)} \\
			\hline
			2 & \(0.819834 - 0.171246\,\mathrm{i}\) & \(0.818254 - 0.170517\,\mathrm{i}\)
			& \(0.770250 - 0.167715\,\mathrm{i}\) & \(0.771259 - 0.166765\,\mathrm{i}\)
			& \(0.608502 - 0.154291\,\mathrm{i}\) & \(0.609087 - 0.153738\,\mathrm{i}\) \\
			3 & \(1.142500 - 0.170695\,\mathrm{i}\) & \(1.146680 - 0.169265\,\mathrm{i}\)
			& \(1.107300 - 0.168903\,\mathrm{i}\) & \(1.110930 - 0.168283\,\mathrm{i}\)
			& \(0.996999 - 0.162716\,\mathrm{i}\) & \(0.999836 - 0.161246\,\mathrm{i}\) \\
			4 & \(1.466140 - 0.170471\,\mathrm{i}\) & \(1.474700 - 0.167398\,\mathrm{i}\)
			& \(1.438840 - 0.169388\,\mathrm{i}\) & \(1.447060 - 0.166338\,\mathrm{i}\)
			& \(1.354720 - 0.165838\,\mathrm{i}\) & \(1.361590 - 0.163194\,\mathrm{i}\) \\
			\hline
		\end{tabular}
		\caption{QNM frequencies for different values of \(l\) obtained using the sixth‑order WKB approximation and the Prony method.}
		\label{tab:4}
	\end{table*}

	\section{Partial Transmission Probabilities}
	\label{sec:4}
	
In a black hole spacetime, a perturbation field propagating outward is scattered by the effective potential barrier generated by the background curvature. The probability that a perturbation mode with frequency \(\omega\), angular quantum number \(l\), and spin \(s\) penetrates this barrier and reaches infinity is called the partial transmission probability, denoted by \(\Gamma_{l}^{(s)}(\omega)\)\cite{Fabbri:1975sa}. Together with the reflection coefficient \(R\), it satisfies the probability conservation relation
\begin{equation}
	\Gamma_{l}^{(s)}(\omega)=|T|^{2}=1-|R|^{2} 
	\label{eq:29}
\end{equation}

~~In this paper we compute only the transmission probability for a single partial wave and focus on its dependence on the parameters. The total absorption cross section or the total greybody factor requires summing over all partial waves and is beyond the scope of the present work.

~~ The partial transmission probability and the quasinormal modes are governed by the same effective potential. The QNMs are discrete complex frequencies obtained under the boundary conditions of purely ingoing waves at the horizon and purely outgoing waves at infinity, whereas the transmission probability corresponds to a scattering problem with an incident plane wave. The two originate from different boundary conditions and describe the influence of the same potential barrier on the propagation of perturbations. Therefore, the analysis of the transmission probability not only complements the QNM results but also provides a way to examine the parameter effects from the perspective of scattering.

~~ The boundary conditions for the scattering problem differ from those used in the QNM calculation. In the scattering problem one must simultaneously consider the incident, reflected, and transmitted waves, i.e., there is a superposition of the incident and reflected waves at infinity, while only a transmitted wave enters the black hole at the horizon. These can be written as:
\begin{equation}
	\psi \sim e^{-i\omega r_{*}} + R\, e^{i\omega r_{*}}, \quad r_{*} \to +\infty 
	\label{eq:30}
\end{equation}
\begin{equation}
	\psi \sim T\, e^{-i\omega r_{*}}, \quad r_{*} \to -\infty 
	\label{eq:31}
\end{equation}
~~where \(R\) and \(T\) are the reflection and transmission coefficients, respectively, satisfying \(|R|^{2}+|T|^{2}=1\).

~~ When the effective potential exhibits a single-peak structure, the transmission coefficient can be calculated using the WKB semi-classical approximation. In the framework of the sixth-order WKB method, the partial transmission probability is given by the following formula~\cite{Konoplya:2011qq}:
\begin{equation}
	\Gamma_{l}^{(s)}(\omega)=\frac{1}{1+e^{2\pi i\Omega}} 
	\label{eq:32}
\end{equation}
\begin{equation}
	\Omega=\frac{i\left(\omega^{2}-V_{0}\right)}{\sqrt{-2V_{0}''}}-\sum_{k=2}^{6}\Lambda_{k}(\omega) 
	\label{eq:33}
\end{equation}
~~Here, \(\Omega\) has exactly the same form as that appearing in the QNM quantization condition~\eqref{eq:23}. This formula automatically guarantees \(0\le \Gamma_{l}^{(s)}\le1\) for all frequencies.

~~ Although the background spacetime is not asymptotically flat, the effective potential decays to zero at infinity; consequently, the transmission probability still approaches unity in the high-frequency limit, just as in the asymptotically flat case.
	
	\begin{figure*}[htbp]
		\centering
		\includegraphics[width=0.3\textwidth]{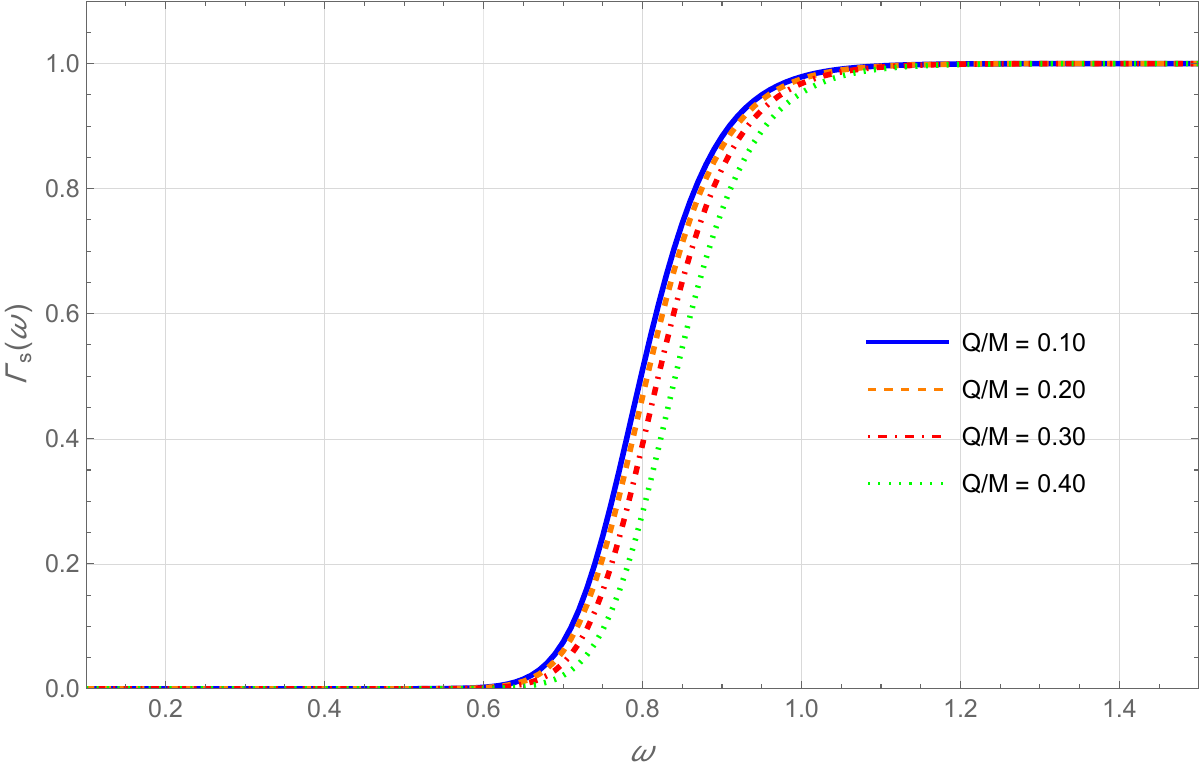}\hfill
		\includegraphics[width=0.3\textwidth]{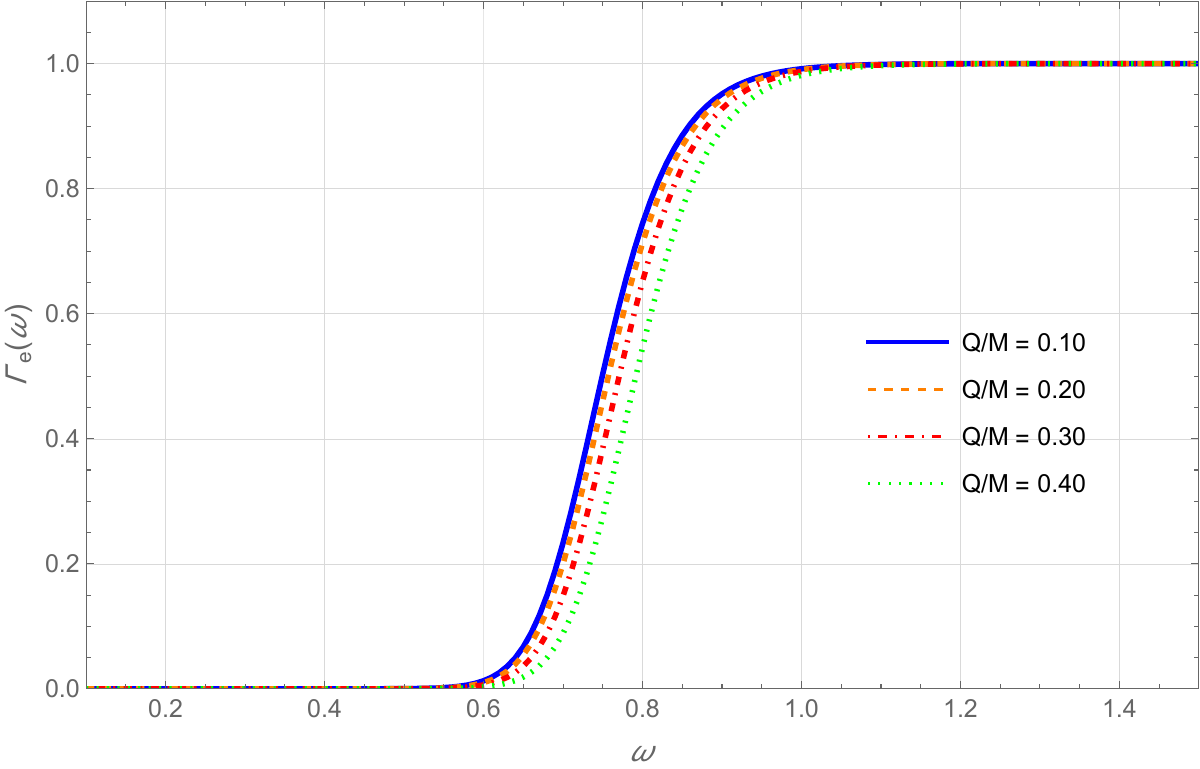}\hfill
		\includegraphics[width=0.3\textwidth]{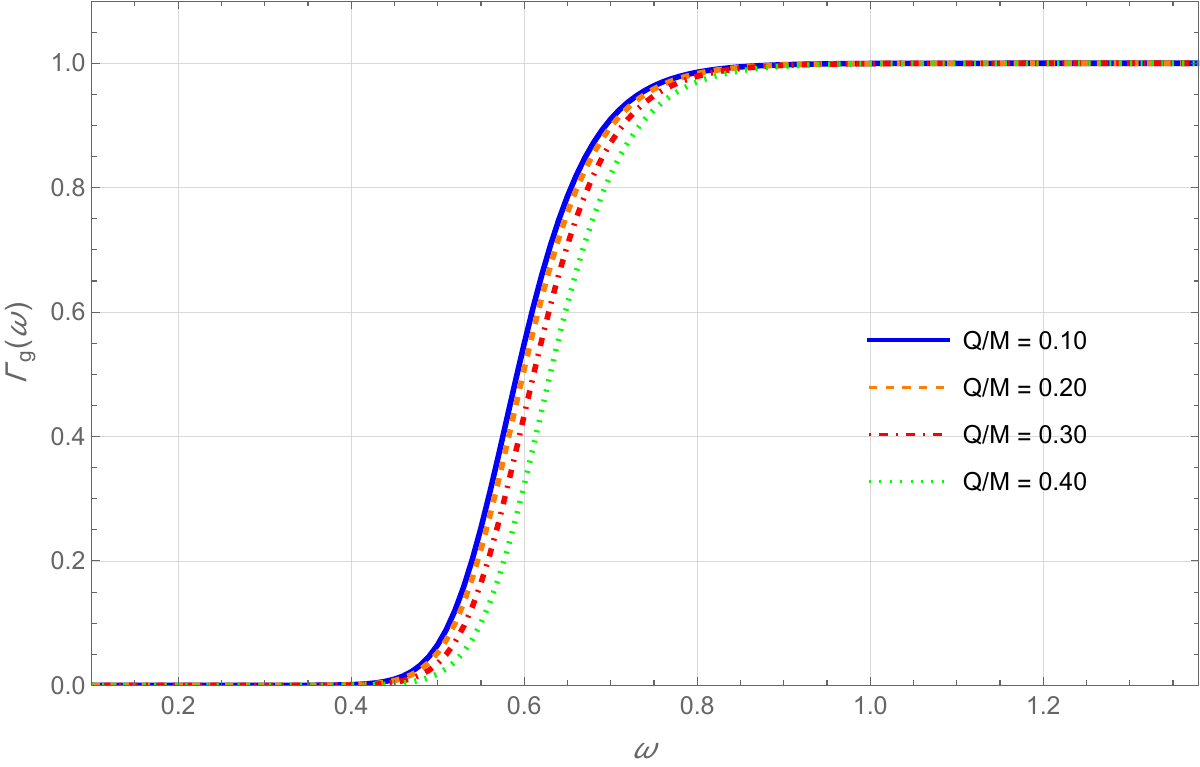}
		\caption{Partial transmission probabilities for the scalar field (left), electromagnetic field (middle), and axial gravitational field (right) for different values of \(Q/M\), with fixed parameters \(l=2\), \(M=1\), \(\tau=0.1\), and \(\lambda/M=0.2\).}
		\label{fig:15}
	\end{figure*}
	
	\begin{figure*}[htbp]
		\centering
		\includegraphics[width=0.3\textwidth]{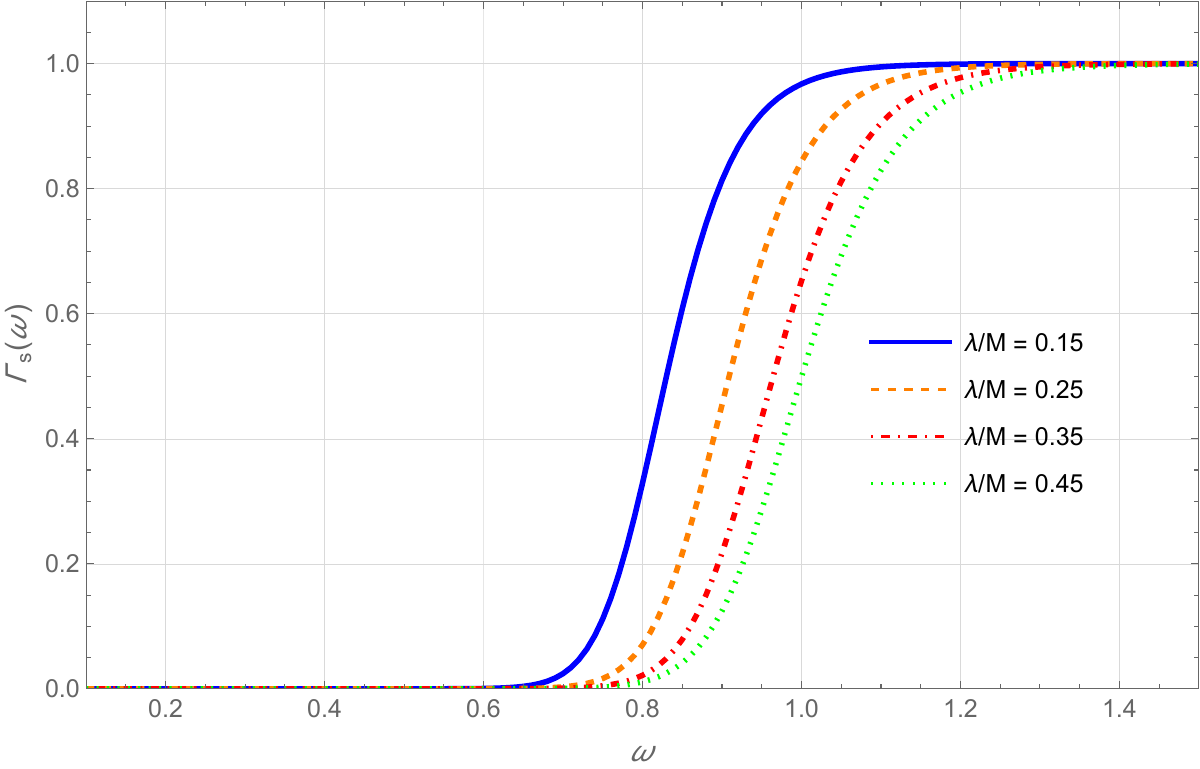}\hfill
		\includegraphics[width=0.3\textwidth]{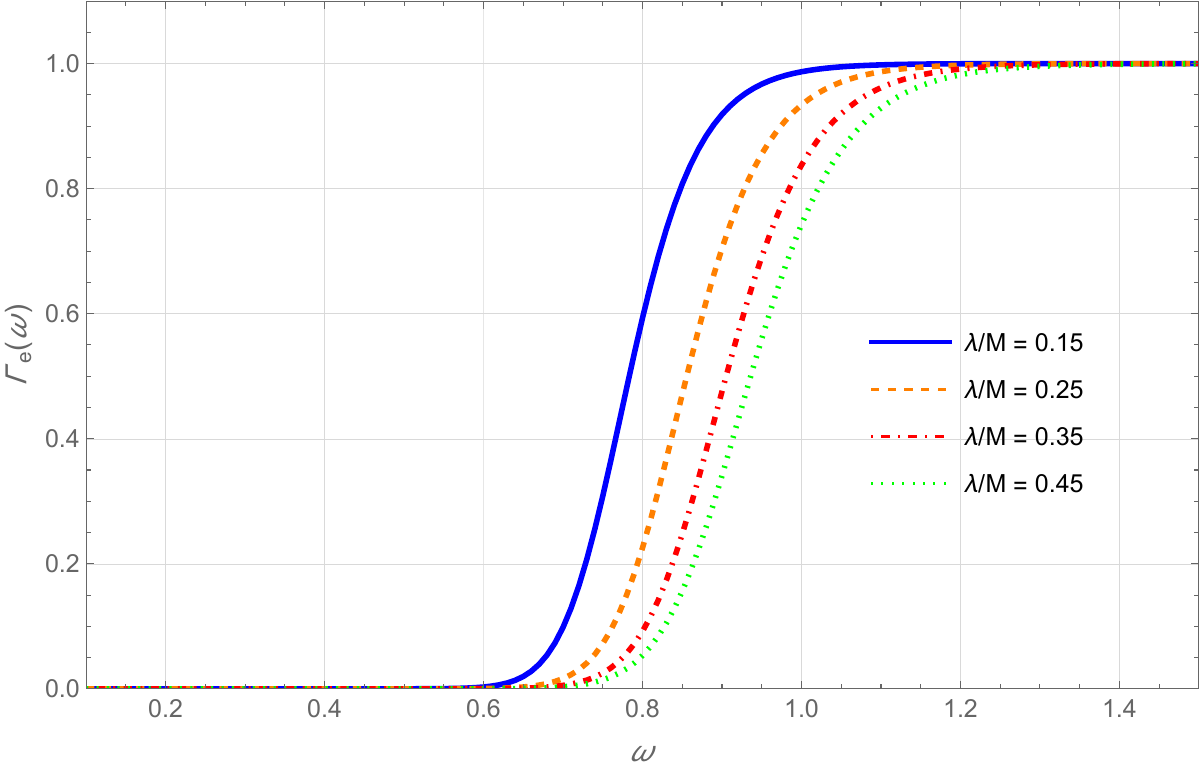}\hfill
		\includegraphics[width=0.3\textwidth]{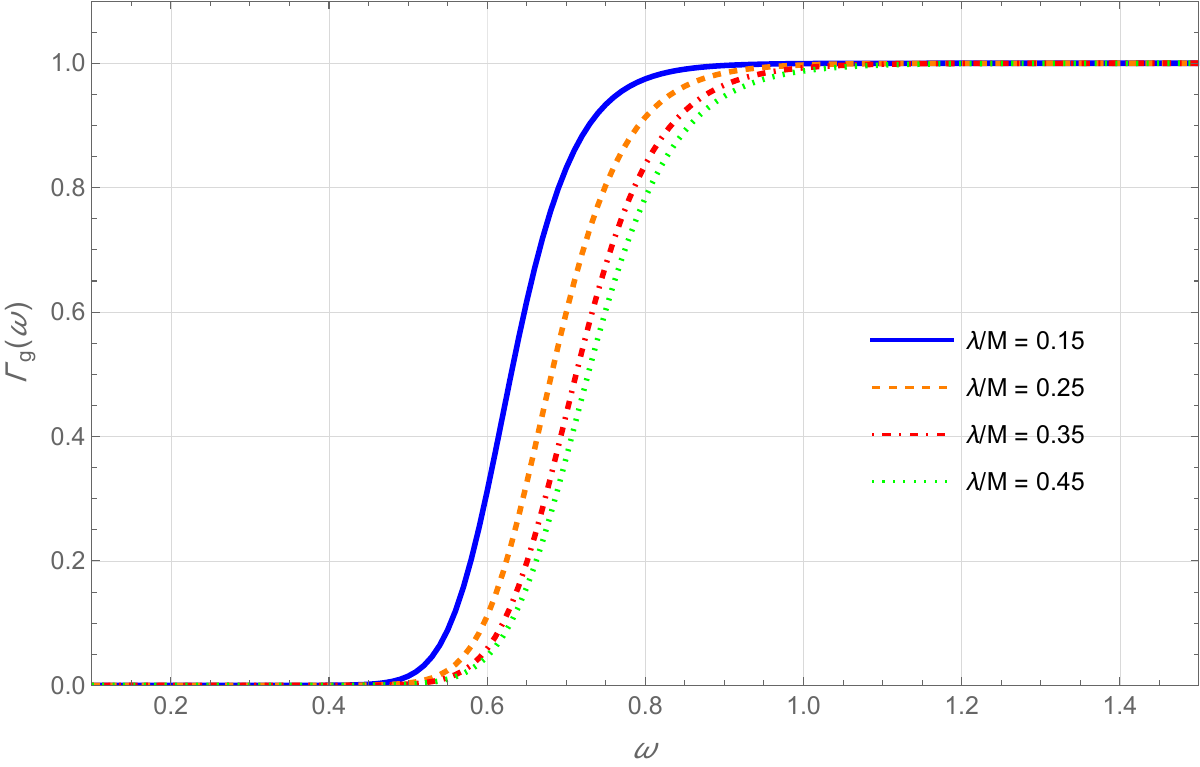}
		\caption{Partial transmission probabilities for the scalar field (left), electromagnetic field (middle), and axial gravitational field (right) for different values of \(\lambda/M\), with fixed parameters \(l=2\), \(M=1\), \(\tau=0.1\), and \(Q/M=0.5\).}
		\label{fig:16}
	\end{figure*}
	
	\begin{figure*}[htbp]
		\centering
		\includegraphics[width=0.3\textwidth]{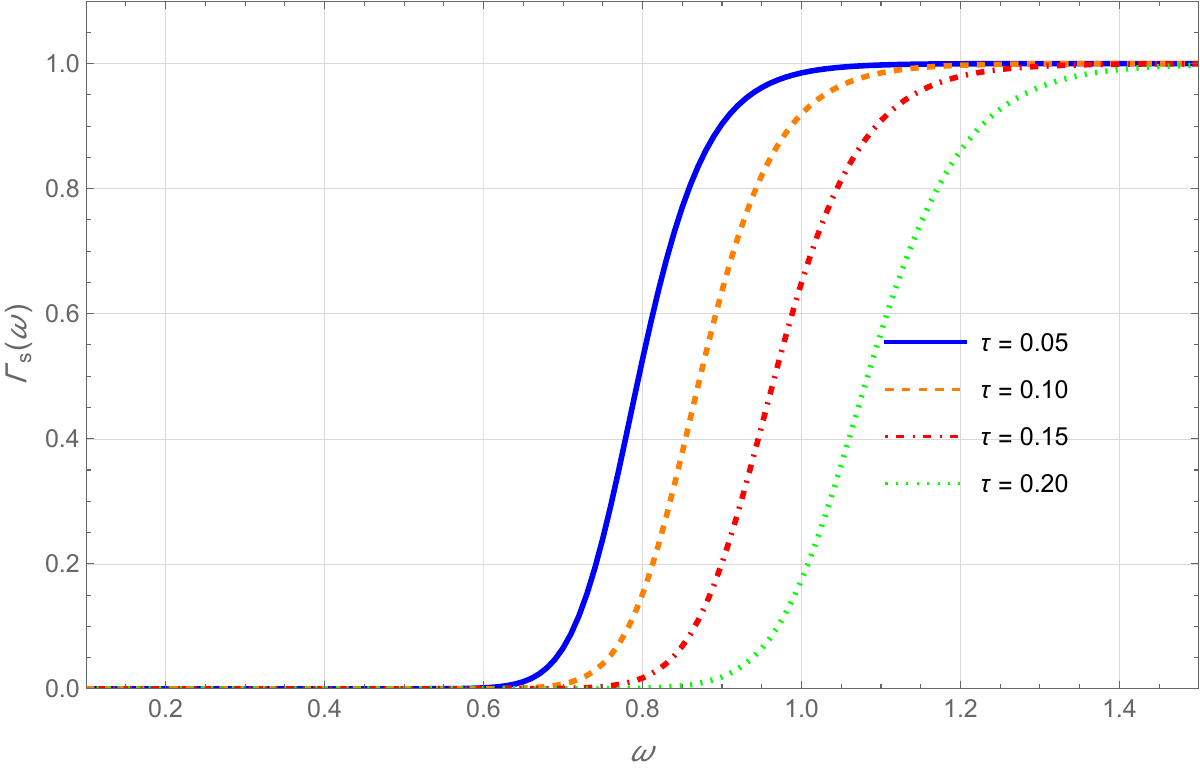}\hfill
		\includegraphics[width=0.3\textwidth]{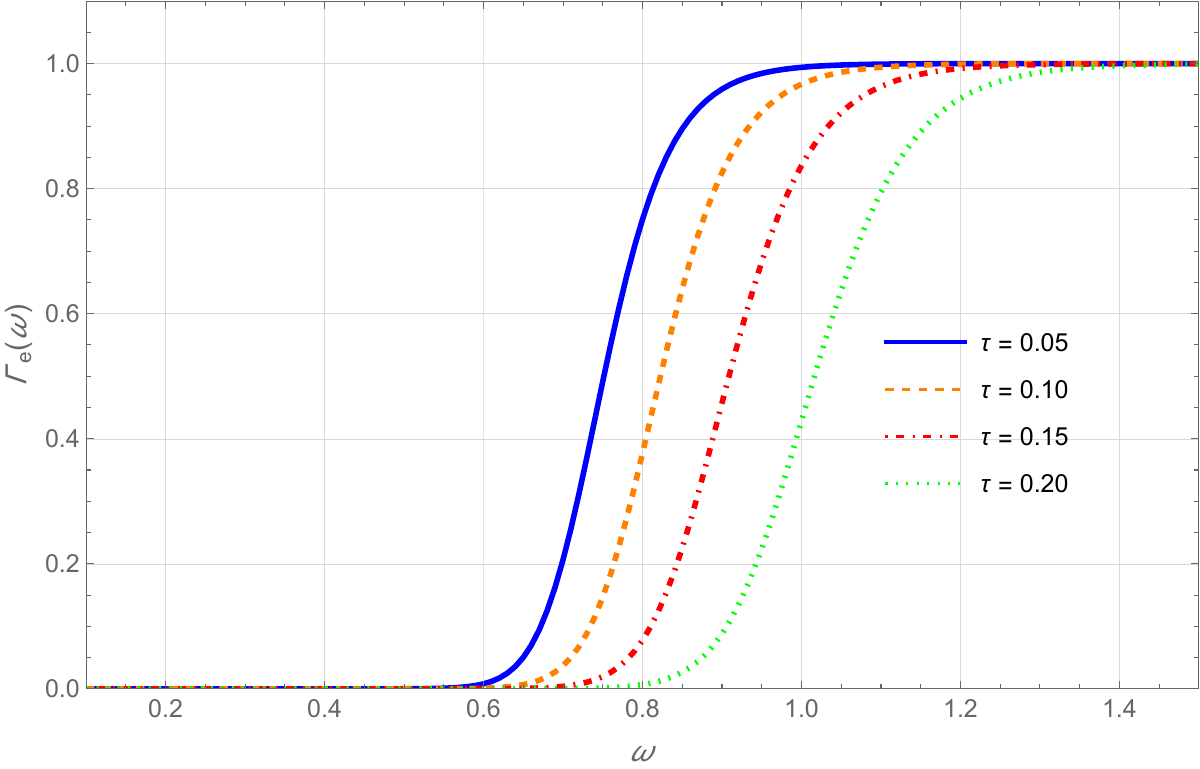}\hfill
		\includegraphics[width=0.3\textwidth]{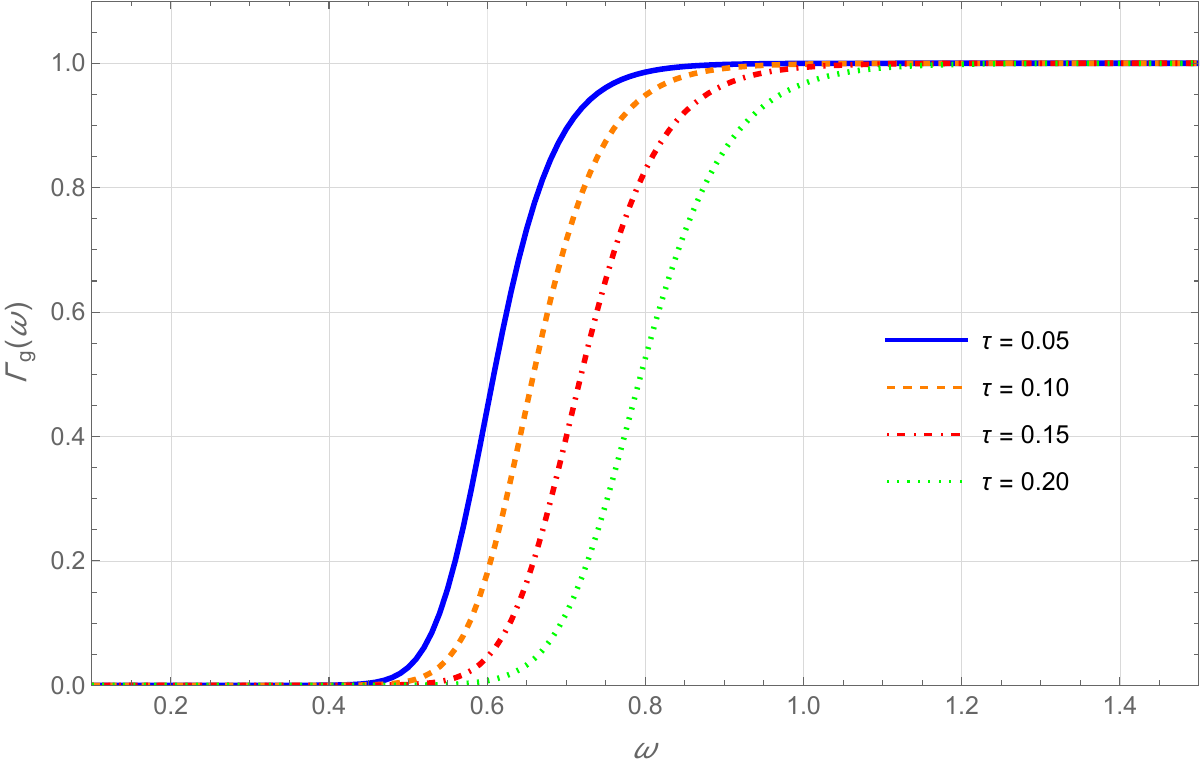}
		\caption{Partial transmission probabilities for the scalar field (left), electromagnetic field (middle), and axial gravitational field (right) for different values of \(\tau\), with fixed parameters \(l=2\), \(M=1\), \(\lambda/M=0.2\), and \(Q/M=0.5\).}
		\label{fig:17}
	\end{figure*}
	
	\begin{figure*}[htbp]
		\centering
		\includegraphics[width=0.3\textwidth]{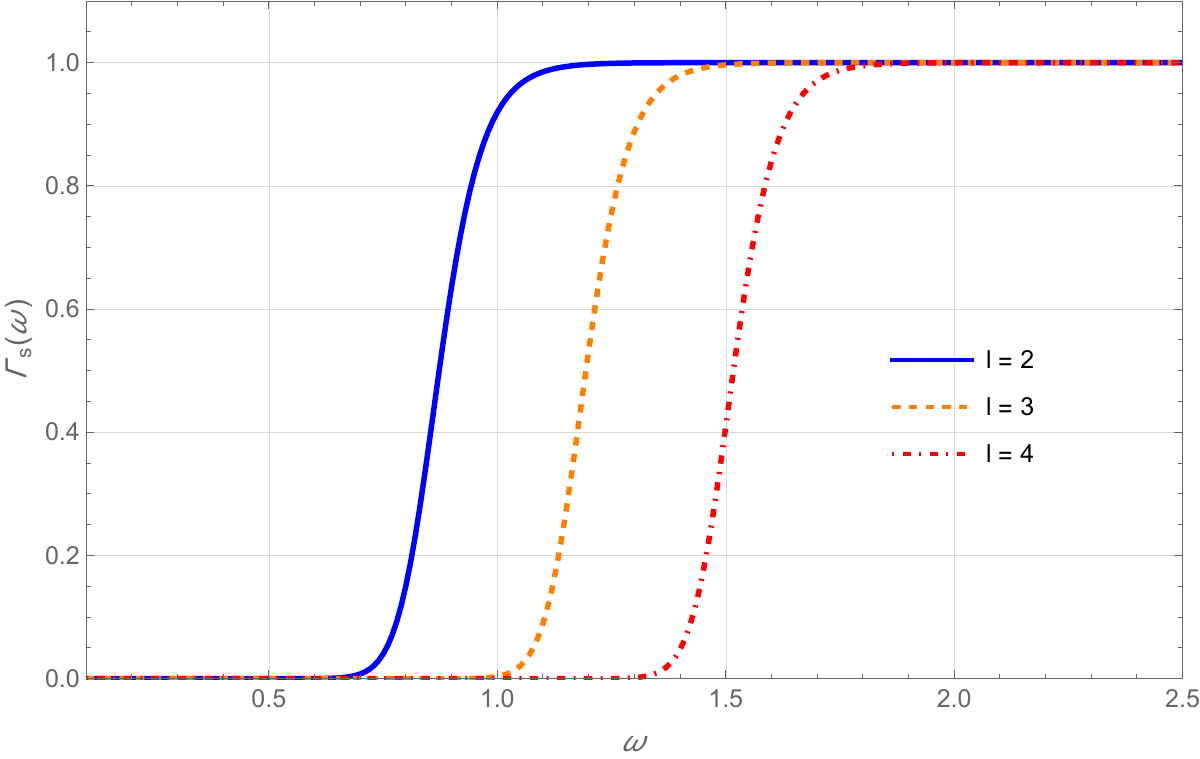}\hfill
		\includegraphics[width=0.3\textwidth]{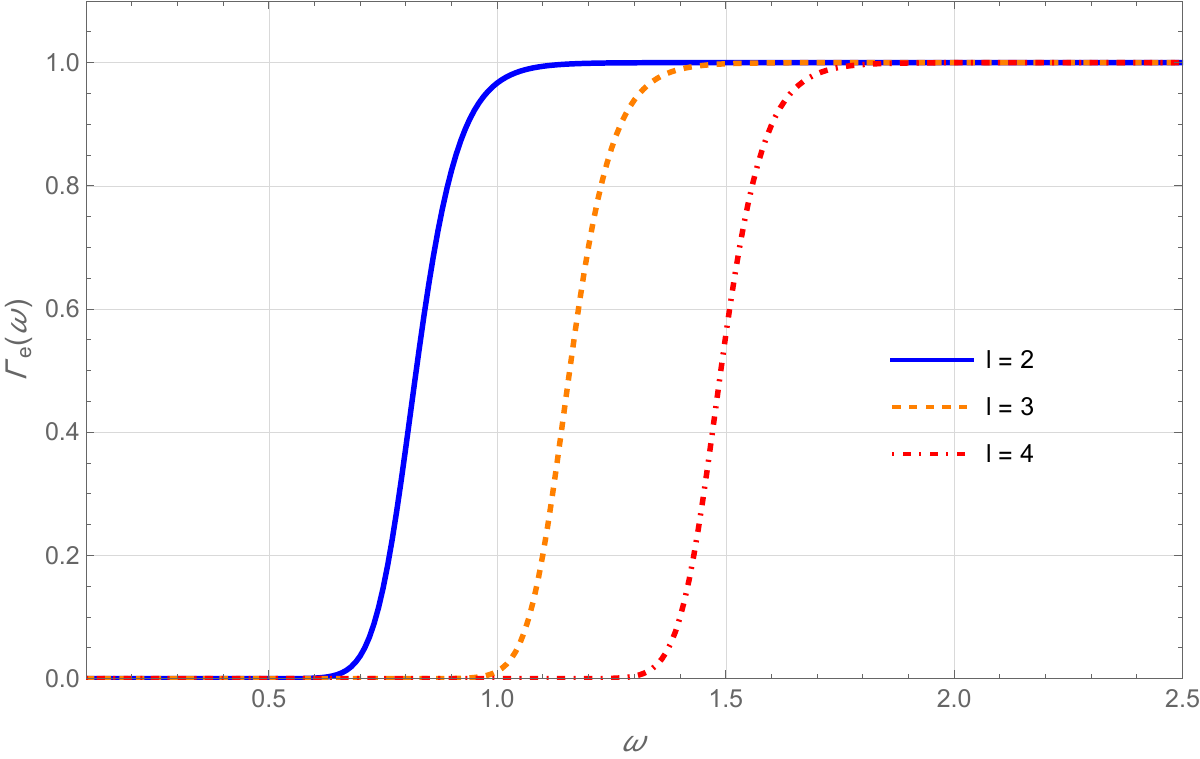}\hfill
		\includegraphics[width=0.3\textwidth]{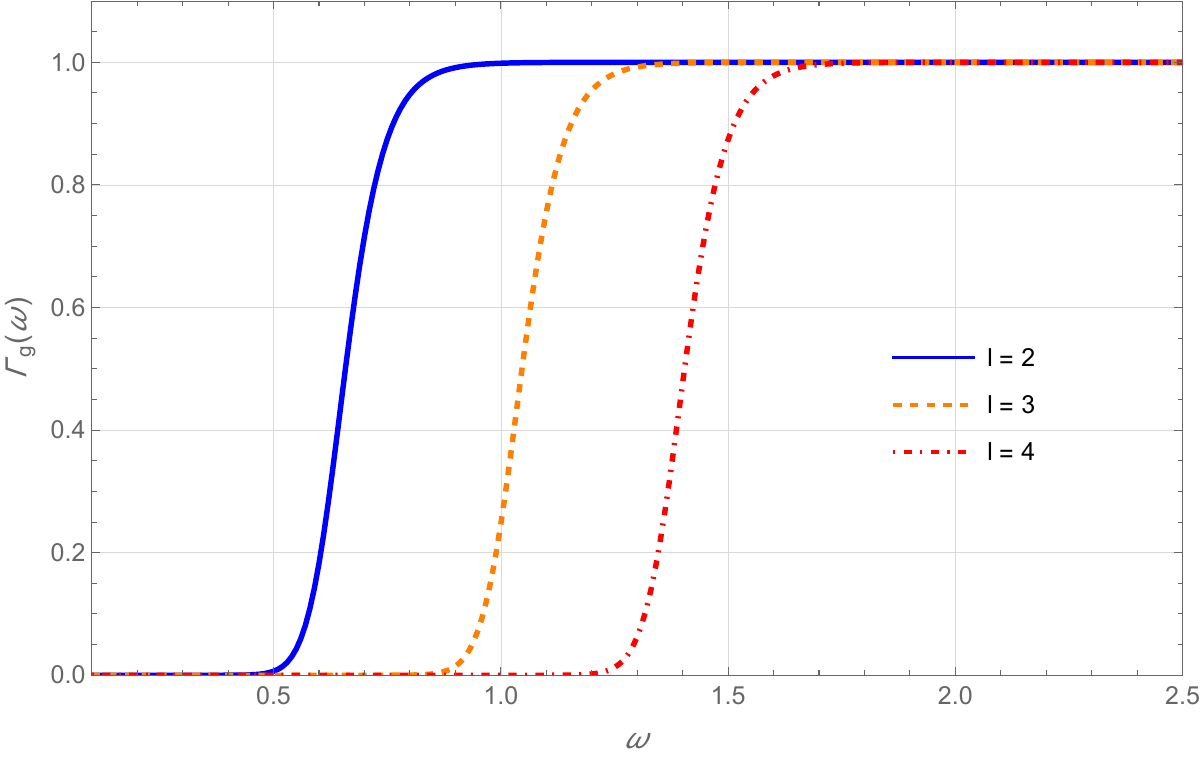}
		\caption{Partial transmission probabilities for the scalar field (left), electromagnetic field (middle), and axial gravitational field (right) for different values of \(l\), with fixed parameters \(M=1\), \(\lambda/M=0.2\), \(Q/M=0.5\), and \(\tau=0.1\).}
		\label{fig:18}
	\end{figure*}
	
	Figures~\ref{fig:15}--\ref{fig:18} display the influence of the charge \(Q/M\), the dark matter parameter \(\lambda/M\), the Lorentz-violating parameter \(\tau\), and the angular quantum number \(l\) on the partial transmission probability for the scalar, electromagnetic, and axial gravitational perturbation, respectively. For all parameter combinations, \(\Gamma_{l}^{(s)}(\omega)\) exhibits two universal features: at low frequencies, \(\Gamma_{l}^{(s)}(\omega)\to0\), indicating that the perturbation does not have enough energy to penetrate the barrier; at high frequencies, \(\Gamma_{l}^{(s)}(\omega)\to1\), meaning that the perturbation energy far exceeds the barrier height and the transmission approaches completion. This overall trend remains unchanged regardless of the parameter values. When a parameter increases, the curve shifts globally toward higher frequencies, which implies that at a given frequency the transmission probability decreases as the parameter increases. At very low frequencies, all curves approach zero and the differences among parameters disappear; at sufficiently high frequencies, all curves tend to unity and the differences likewise vanish. Hence, the influence of the parameters manifests itself only in the intermediate frequency range. Under the same parameters, the transmission probabilities of the three perturbation always satisfy \(\Gamma_{l}^{(2)}>\Gamma_{l}^{(1)}>\Gamma_{l}^{(0)}\), i.e., the gravitational field has the highest transmission probability, followed by the electromagnetic field, and the scalar field has the lowest. This ordering corresponds to the ordering of the effective potential heights \(V_{g}<V_{em}<V_{s}\), namely, the lower the barrier, the easier the tunneling and the higher the transmission probability. 
	
	~~Figure~\ref{fig:15} shows that the charge \(Q/M\) has the weakest influence on the transmission probability among the three parameters; the curve shifts only slightly toward higher frequencies, which is consistent with the earlier effective potential analysis where the charge term only mildly elevates the barrier. In Fig.~\ref{fig:16}, the dark matter parameter \(\lambda/M\) exerts a markedly stronger influence than \(Q/M\) on the transmission probability, and the curve shifts more significantly. This originates from the fact that the dark matter term raises the effective potential barrier over a wide radial range. In Fig.~\ref{fig:17}, the Lorentz-violating parameter \(\tau\) has the most prominent influence among the three parameters; the curve is substantially shifted toward higher frequencies. At \(\tau=0.20\), the transmission probability of the scalar perturbation is almost zero within the range \(\omega\lesssim0.8\). When the parameter \(\tau\) increases, the shift of the gravitational perturbation curve is slightly smaller than those of the scalar and electromagnetic perturbations, which echoes the behavior observed in the QNM analysis where the gravitational perturbation frequency increases more slowly. In Fig.~\ref{fig:18}, increasing the angular quantum number \(l\) likewise shifts the curve toward higher frequencies, and the transition becomes more rapid. The mechanism behind this is the raising of the barrier height through the centrifugal-dominated term.
	
	\section{Validity of Hod’s conjecture}
	\label{sec:5}
	
	Hod's conjecture states that for any black hole, the absolute value of the imaginary part of the QNM frequency should be bounded from above by the Hawking temperature~\cite{Hod:2006jw,Gogoi:2024epx}:
	\begin{equation}
		\left|\Im(\omega)\right| \le \pi T_H
		\label{eq:34}
	\end{equation}
	~~where \(T_H\) is the Hawking temperature of the black hole. The physical significance of this conjecture is profound: it indicates that the decay rate of black hole perturbations cannot exceed a universal upper limit set by the Hawking temperature---that is, the dissipation rate of a black hole cannot be ``too fast''. This inequality connects the classical perturbation theory of black holes with their quantum thermodynamic properties, suggesting a deep relation between black hole stability and thermodynamic entropy.
	
	For the charged KR--PFDM black hole studied in this paper, the spacetime is not asymptotically flat; therefore the calculation of the Hawking temperature requires an appropriate normalization of the Killing vector at infinity. Using the surface gravity method, the Hawking temperature is derived as~\cite{Ahmed:2026doo}
	\begin{equation}
		T_H = \frac{\sqrt{1-\tau}}{4\pi r_h}\left[\frac{1}{1-\tau} - \frac{Q^2}{(1-\tau)^2 r_h^{2}} + \frac{\lambda}{r_h}\right] \label{eq:35}
	\end{equation}
	~~where \(r_h\) is the event horizon radius. This expression reduces to the standard Schwarzschild result in the limits $Q\to0$, \(\tau\to0\) and \(\lambda\to0\). 
	
		\begin{figure*}[htbp]
		\centering
		\renewcommand{\thesubfigure}{\roman{subfigure}}  
		\subcaptionbox{$\tau=0.1$, $\lambda/M=0.2$}{
			\includegraphics[width=0.3\textwidth]{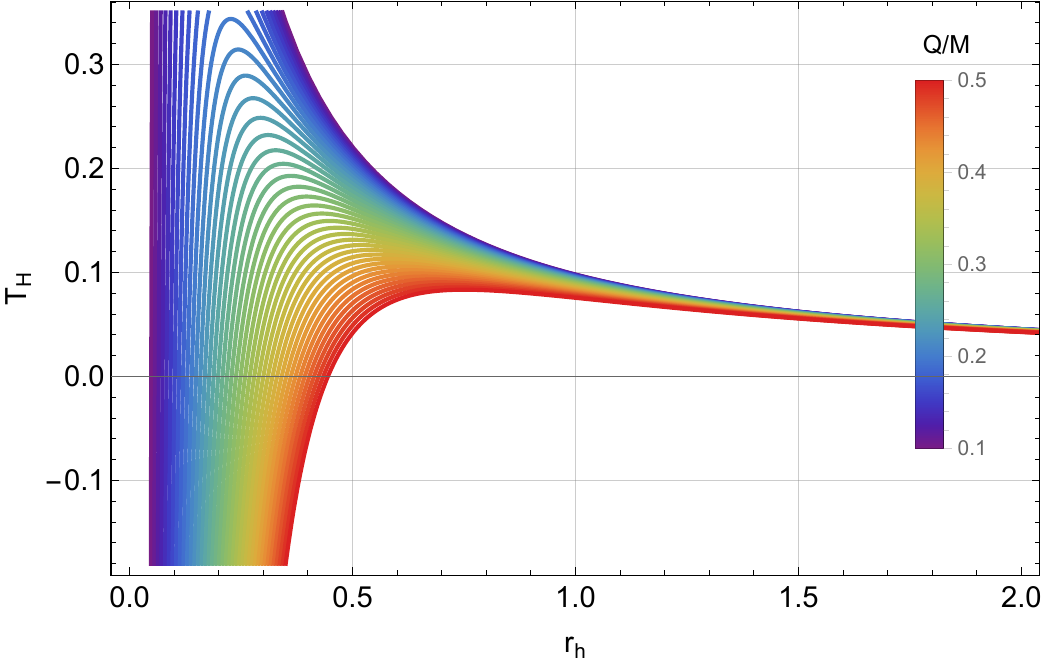}
		}
		\hfill
		\subcaptionbox{$\tau=0.1$, $Q/M=0.5$}{
			\includegraphics[width=0.3\textwidth]{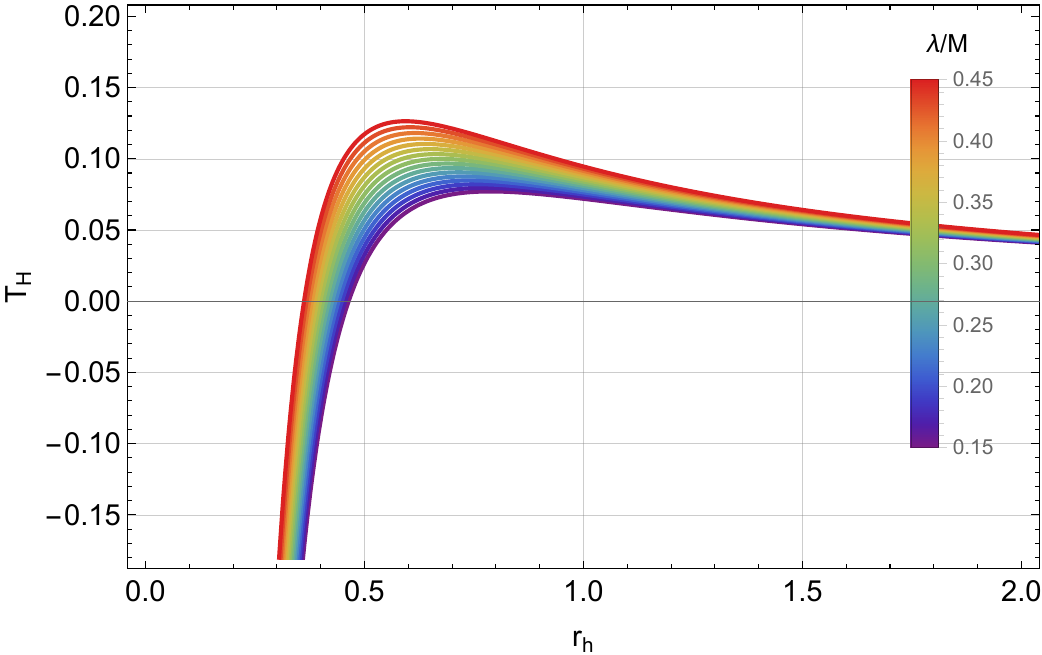}
		}
		\hfill
		\subcaptionbox{$\lambda/M=0.2$, $Q/M=0.5$}{
			\includegraphics[width=0.3\textwidth]{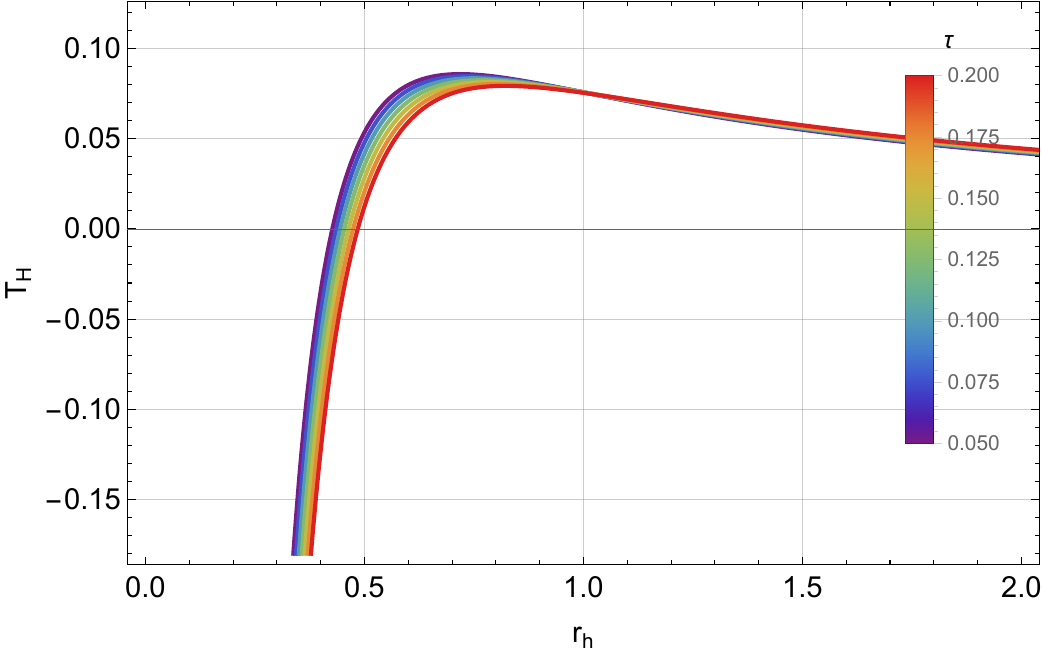}
		}
		\caption{Variation of the Hawking temperature \(T_H\) with the event horizon radius \(r_h\) for different values of the parameters.}
		\label{fig:19}
	\end{figure*}
	
	Figure~\ref{fig:19} displays the variation of the Hawking temperature \(T_H\) with the event horizon radius \(r_h\). Each curve exhibits a single‑peak structure, rising first and then declining. At the small‑radius end, the temperature tends to zero or drops to negative values, corresponding to the vanishing of the surface gravity in the extremal black hole limit. At the large‑radius end, the temperature decays as \(T_H = \frac{1}{4\pi\sqrt{1-\tau}\,r_h} + \frac{\lambda\sqrt{1-\tau}}{4\pi r_h^2} + \mathcal{O}(r_h^{-3})\), where the first term is the leading contribution, scaling as \(1/r_h\), with its coefficient modified by the parameter \(\tau\); the second term is the subleading term induced by dark matter. The three panels show that as the charge \(Q/M\) or the Lorentz‑violating parameter \(\tau\) increases, the \(T_H\) curve shifts downward and the temperature decreases, whereas as the dark matter parameter \(\lambda/M\) increases, the \(T_H\) curve shifts upward and the temperature rises. This difference originates from the fact that, in the expression for the Hawking temperature, the charge term and the \(\tau\)‑related correction terms act to reduce the temperature, while the dark matter term acts to increase it.
	
\begin{figure*}[htbp]
		\centering
		\includegraphics[width=0.30\textwidth]{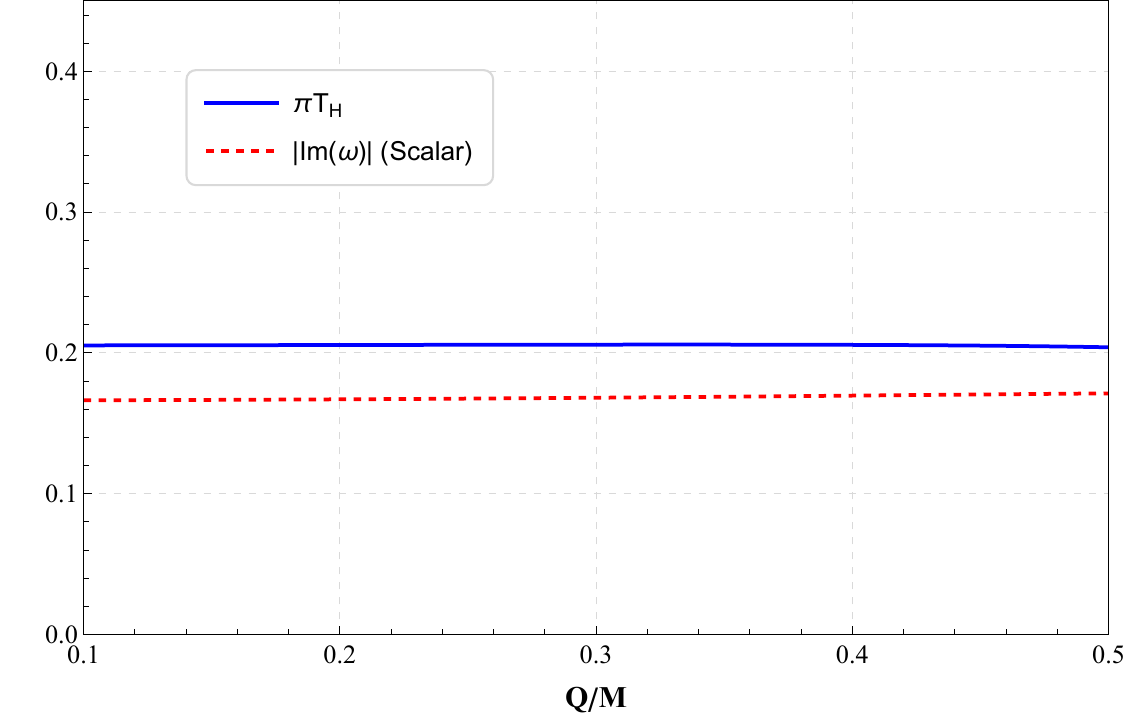}\hfill
		\includegraphics[width=0.30\textwidth]{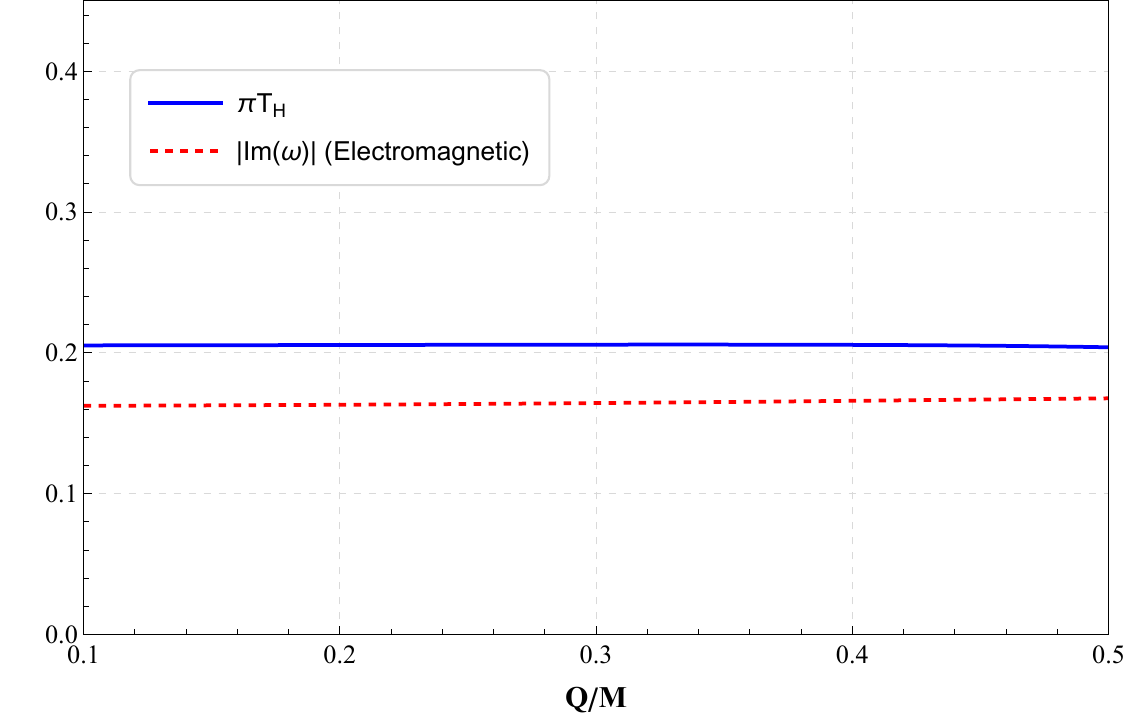}\hfill
		\includegraphics[width=0.30\textwidth]{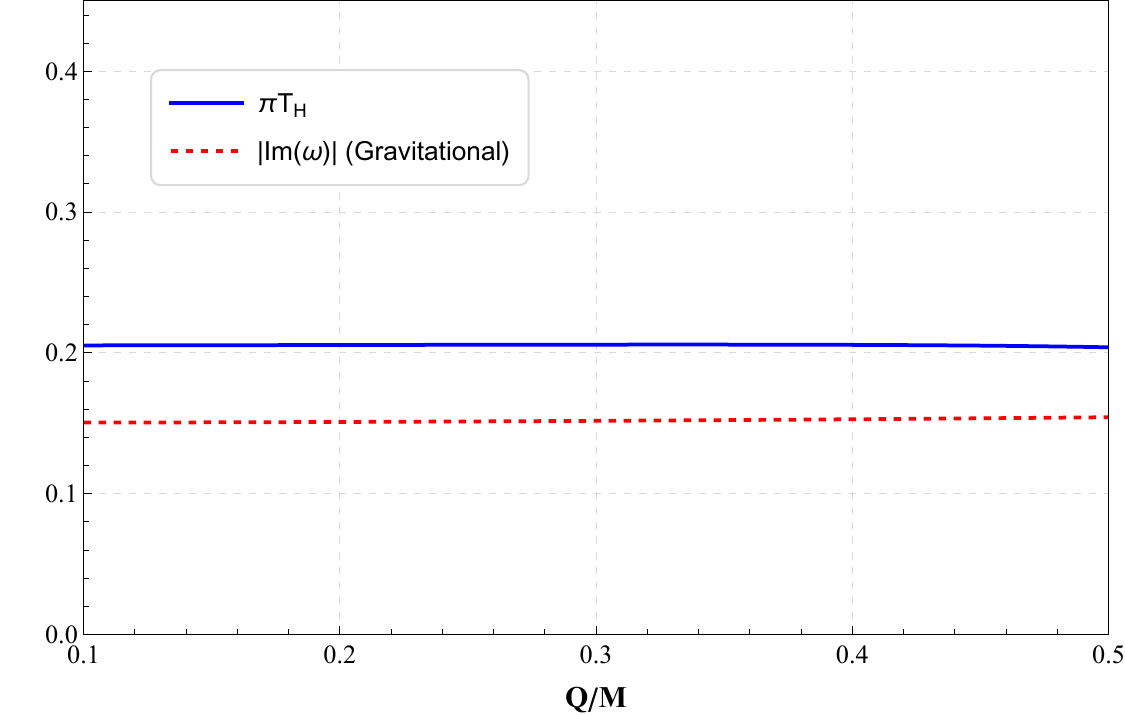}
		\caption{Variation of \(|\Im(\omega)|\) and \(\pi T_H\) with \(Q/M\) for different perturbation fields, with fixed parameters \(M=1\), \(l=2\), \(\tau=0.1\), and \(\lambda/M=0.2\).}
		\label{fig:20}
	\end{figure*}

\begin{figure*}[!htbp]
		\centering
		\includegraphics[width=0.3\textwidth]{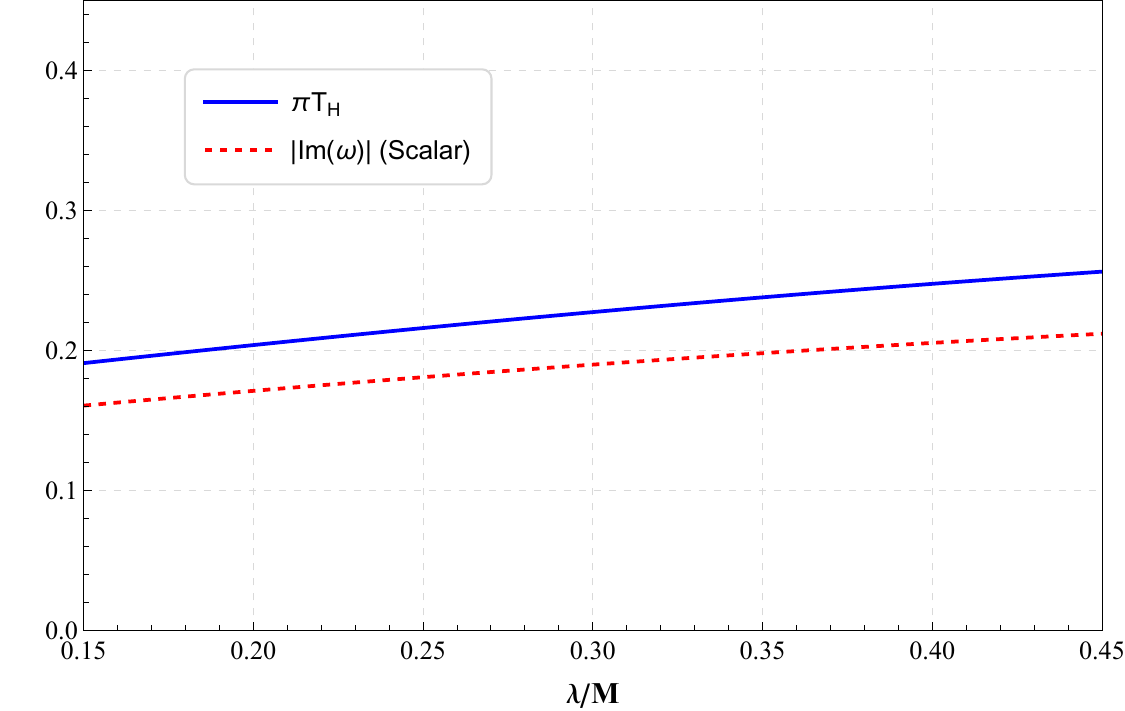}\hfill
		\includegraphics[width=0.3\textwidth]{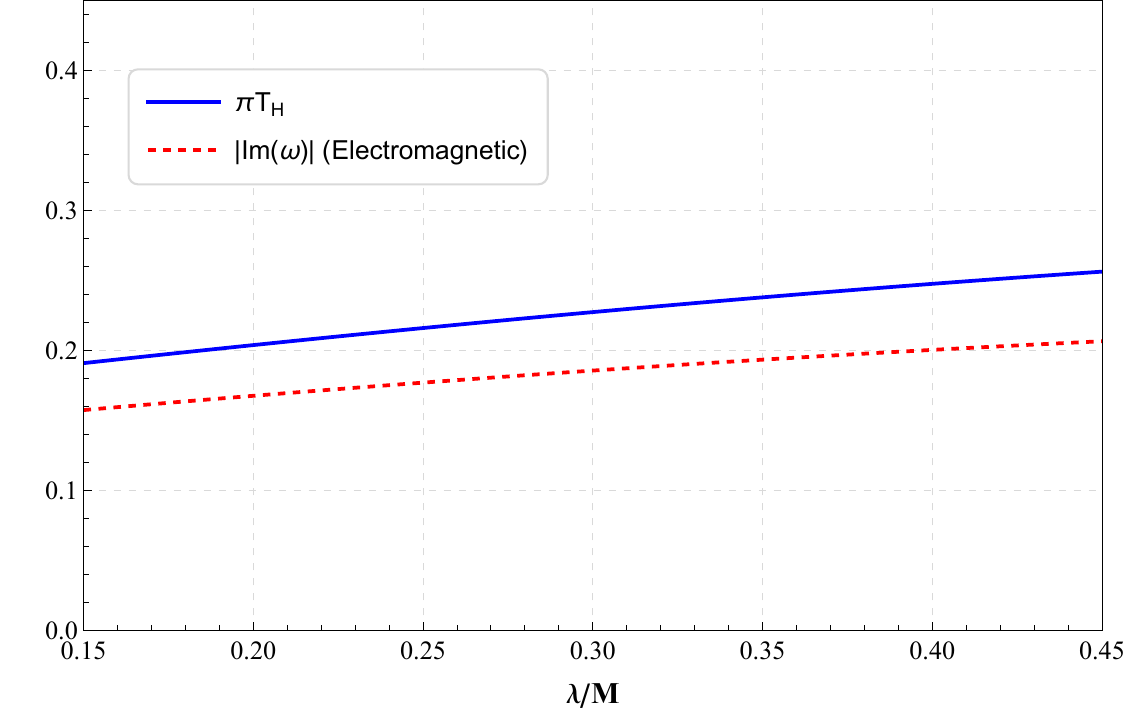}\hfill
		\includegraphics[width=0.3\textwidth]{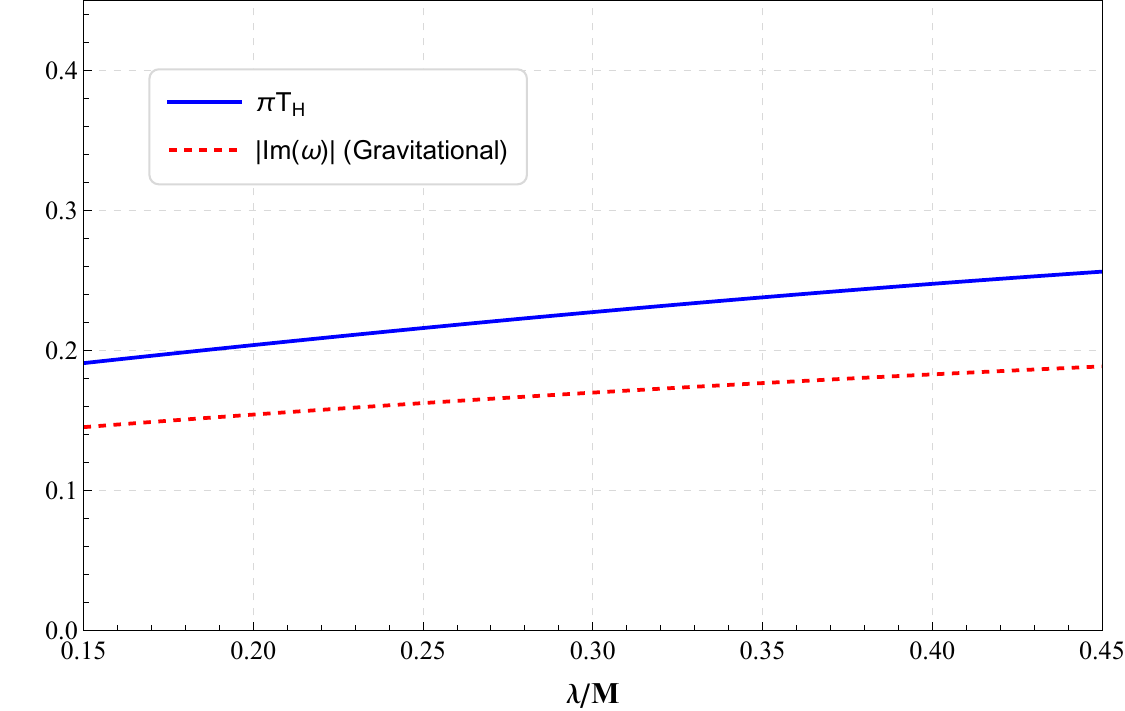}
		\caption{Variation of \(|\Im(\omega)|\) and \(\pi T_H\) with \(\lambda/M\) for different perturbation fields, with fixed parameters \(M=1\), \(l=2\), \(\tau=0.1\), and \(Q/M=0.5\).}
		\label{fig:21}
	\end{figure*}

\begin{figure*}[!htbp]
		\centering
		\includegraphics[width=0.3\textwidth]{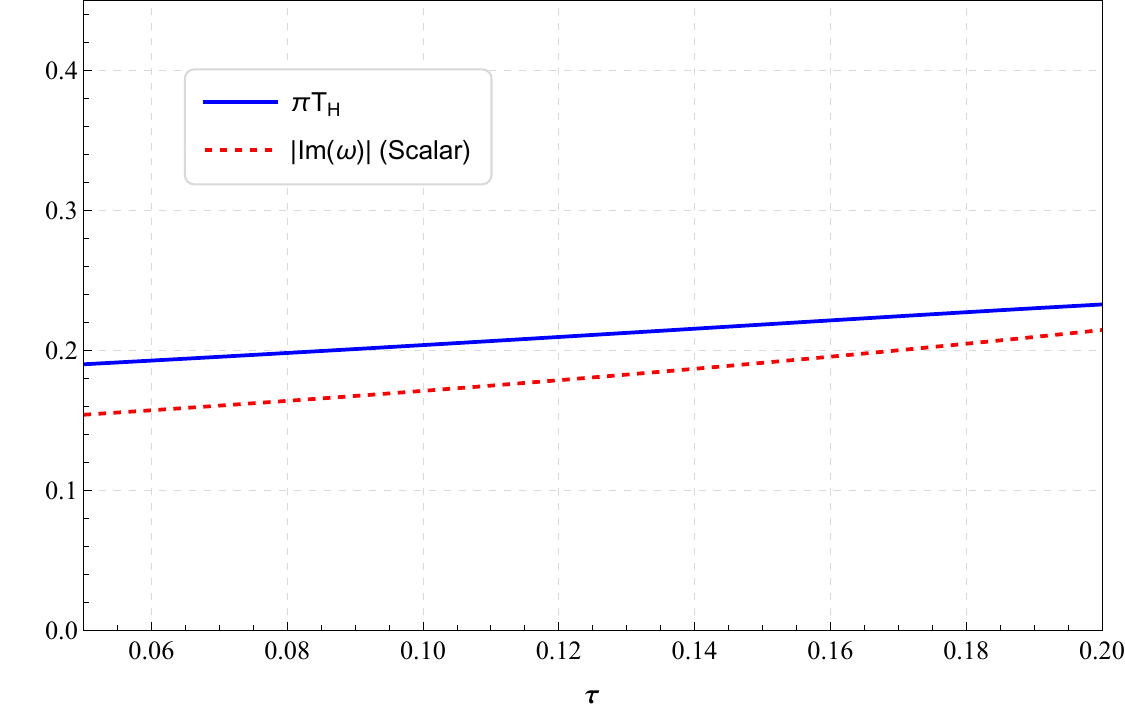}\hfill
		\includegraphics[width=0.3\textwidth]{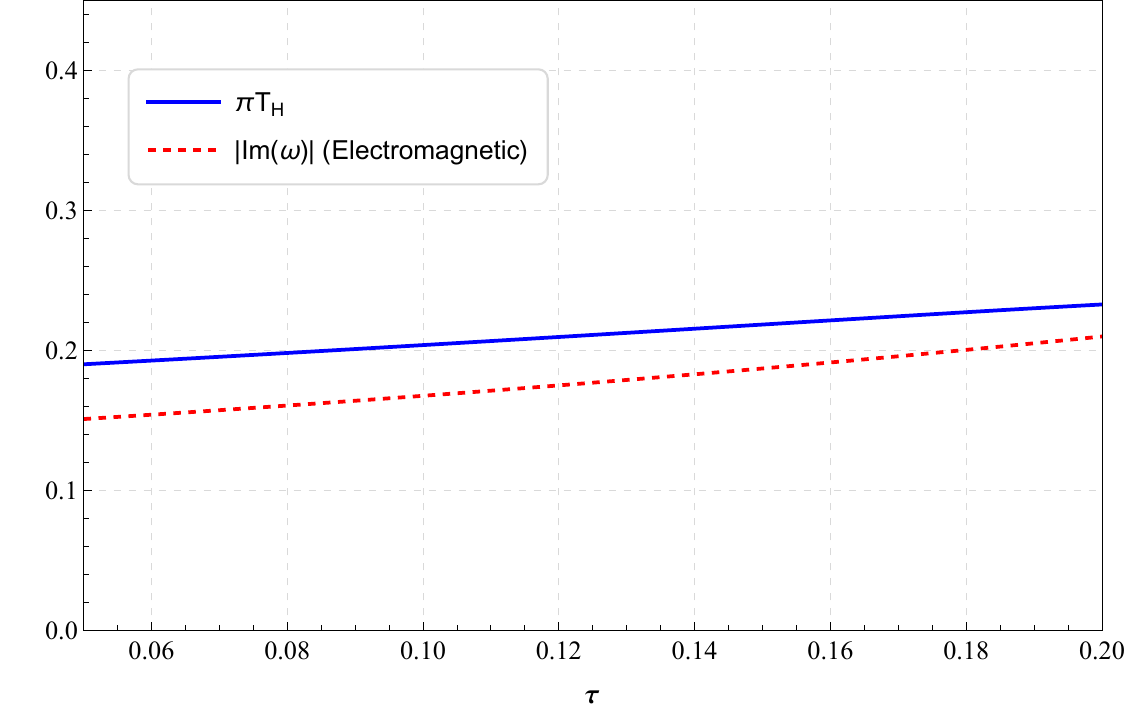}\hfill
		\includegraphics[width=0.3\textwidth]{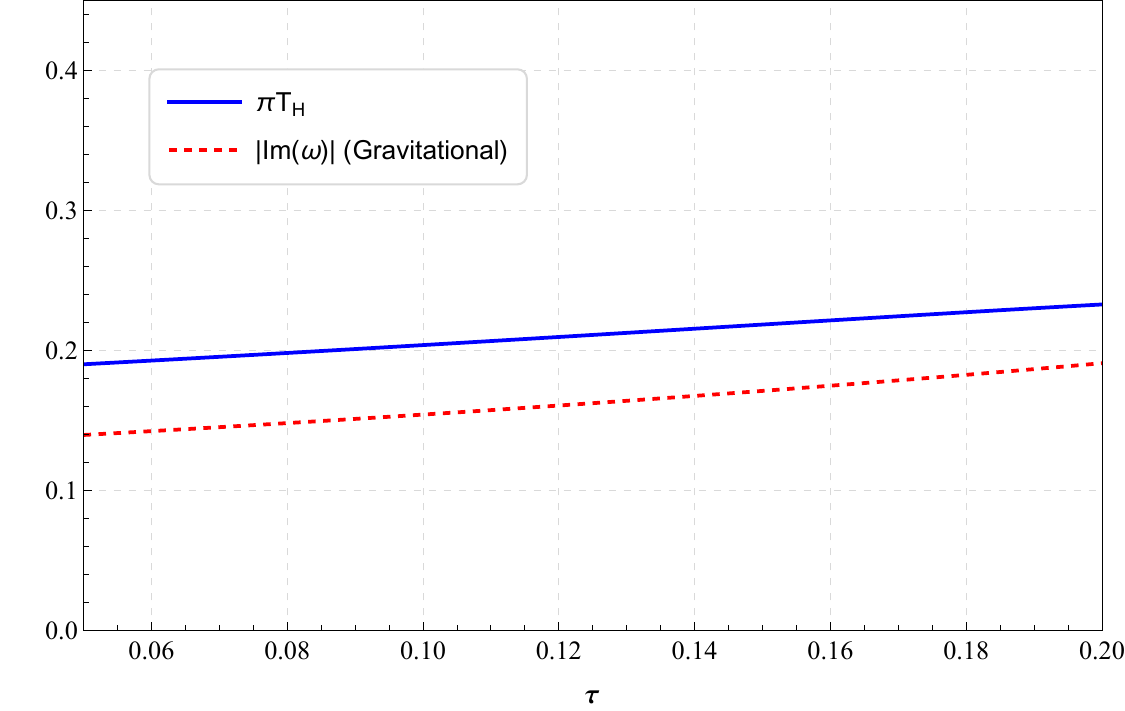}
		\caption{Variation of \(|\Im(\omega)|\) and \(\pi T_H\) with \(\tau\) for different perturbation fields, with fixed parameters \(M=1\), \(l=2\), \(\lambda/M=0.2\), and \(Q/M=0.5\).}
		\label{fig:22}
	\end{figure*}

	Figs.~\ref{fig:20} ,\ref{fig:21} and \ref{fig:22} present the variation of \(\left|\Im(\omega)\right|\) and \(\pi T_H\) with each of the parameters for the three types of perturbations. In all calculations, \(\left|\Im(\omega)\right|\) always lies below the \(\pi T_H\) curve, indicating that Hod's conjecture holds over the entire parameter range examined. The relative change of the two curves differs among the parameters. When \(Q/M\) increases, \(\left|\Im(\omega)\right|\) rises slightly while \(\pi T_H\) remains nearly unchanged, so the gap between them gradually narrows. This occurs because the charge term mainly influences the QNM damping rate by modifying the effective potential, whereas its effect on the Hawking temperature is relatively weak. When \(\lambda/M\) increases, both \(\left|\Im(\omega)\right|\) and \(\pi T_H\) increase, but the rise of \(\pi T_H\) is more pronounced, and the gap widens somewhat. The physical reason is that the dark matter term enters both the Hawking temperature expression and the effective potential, and its enhancing effect on the temperature is stronger than that on the damping rate. When \(\tau\) increases, both \(\left|\Im(\omega)\right|\) and \(\pi T_H\) increase significantly, but the increase of \(\left|\Im(\omega)\right|\) is larger, so the gap rapidly shrinks. The parameter \(\tau\) modifies the constant term and the charge term of the metric function as a factor; the barrier-raising effect is mainly reflected in the rapid growth of \(\left|\Im(\omega)\right|\), while its impact on the Hawking temperature is partly offset by the factor \(\sqrt{1-\tau}\) in the temperature formula.
	

	\section{Summary}
	\label{sec:6}
	
	Within the KR field-induced Lorentz-violating gravity, we have investigated the perturbation dynamics and QNM spectra of a charged black hole immersed in PFDM. The spacetime is not asymptotically flat, with the asymptotic value of the metric function given by \(f(\infty)=1/(1-\tau)\); this non-asymptotic flatness is a direct consequence of the KR field modifying the spacetime geometry. 
	
	~~Under the test-field approximation, the equations of motion for three types of perturbations (scalar, electromagnetic, and axial gravitational fields) can all be reduced to Schr\"{o}dinger-type radial equations, and the effective potentials exhibit a single-peak structure. The peak height of the effective potential increases monotonically with each parameter, while the shift direction of the peak in the tortoise coordinate depends on the type of parameter and the spin of the perturbation field. The angular quantum number \(l\) raises the barrier height through the centrifugal-dominated term, causing a significant increase in the oscillation frequency while the damping rate changes very little. Under the same parameters, both the effective potential heights and the QNM frequencies satisfy the ordering ``scalar field $>$ electromagnetic field $>$ gravitational field''. 
	
	~~The QNM frequencies are computed using the sixth-order WKB method in the frequency domain and the Prony method in the time domain, and the results obtained from the two methods are in excellent agreement. Parameter scans show that \(\tau\) has the strongest influence on the QNM frequencies, followed by \(\lambda/M\), while \(Q/M\) has the weakest effect. The results for the partial transmission probabilities indicate that increasing any of the parameters shifts the curves globally toward higher frequencies and reduces the transmission probability, which is consistent with the trend of the QNM damping rates. The two describe the influence of the same potential barrier on the propagation of perturbations, respectively from the perspectives of scattering and resonant states. In the test of Hod's conjecture, \(\left|\Im(\omega)\right|\) always satisfies \(\left|\Im(\omega)\right| \le \pi T_H\) within the investigated parameter range. The relative variation of the two curves differs among the parameters: the gap between them narrows as \(Q/M\) or \(\tau\) increases, whereas it widens as \(\lambda/M\) increases. 
	
	~~In summary, Lorentz violation, electric charge, and dark matter all affect the perturbation propagation and QNM spectra of black holes, with different strengths and mechanisms. In particular, the Lorentz-violating parameter \(\tau\) has a much stronger influence on the QNM frequencies than the charge and dark matter parameters, providing a potential discriminatory signature for distinguishing the effects of Lorentz violation from dark matter in future gravitational wave observations.
	
	\section*{Acknowledgments}
	This work was supported by the National Natural Science Foundation of China (No.12265007), and the Guizhou Provincial Major Scientific and Technological Program (XKBF(2025)010).

	\newpage
	
	
	\bibliographystyle{unsrt}  
	\bibliography{ref1}
	\bibliographystyle{apsrev4-1}

\end{document}